\documentclass[11pt,a4paper]{article}
\pdfoutput=1  
\usepackage{epsfig}
\usepackage{graphicx,psfrag}
\usepackage{multirow}
\usepackage{slashed}
\usepackage{pstricks}
\usepackage{caption}
\usepackage{subcaption}
\usepackage{url}
\usepackage{braket}
\usepackage{jheppub}
\usepackage{enumerate}
\usepackage{wasysym} 
\usepackage{mathrsfs} 
\usepackage{amsfonts} 
\usepackage{amsbsy} 
\usepackage{amscd}
\usepackage{color}
\usepackage{changepage}
\usepackage{floatrow}

\makeatletter

\def\eeq{\end{equation}}
\def\beq{\begin{equation}}

\newcommand{\Rmnum}[1]{\expandafter\@slowromancap\romannumeral #1@}

\newcommand{\bea} {\begin{eqnarray}}
\newcommand{\eea} {\end{eqnarray}}

\newcommand{\gsim}{\raisebox{-0.13cm}{~\shortstack{$>$ \\[-0.07cm]
      $\sim$}}~}
\newcommand{\lsim}{\raisebox{-0.13cm}{~\shortstack{$<$ \\[-0.07cm]
      $\sim$}}~}
\makeatother

\title{High scale validity of two Higgs doublet scenarios with a real scalar singlet dark matter }
\author[a]{Subhaditya Bhattacharya,}
   \affiliation[a]{Department of Physics, Indian Institute of Technology Guwahati,
North Guwahati, Assam - 781039, India}

\author[b]{Atri Dey,}
  \affiliation[b]{School of Theoretical Physics, Dublin Institute for Advanced Studies,
10 Burlington Road, Dublin 4, Ireland}

\author[c]{Jayita Lahiri,}
   \affiliation[c]{II. Institut f{\"u}r Theoretische Physik, Universit{\"a}t Hamburg, Luruper Chaussee 149, 22761 Hamburg, Germany}

\author[d]{Biswarup Mukhopadhyaya} 
  \affiliation[d]{Department of Physical Sciences, Indian Institute of Science Education and Research Kolkata, Mohanpur - 741246, India}

\emailAdd{subhab@iitg.ac.in}
\emailAdd{atri@stp.dias.ie}
\emailAdd{jayita.lahiri@desy.de}
\emailAdd{biswarup@iiserkol.ac.in}

\abstract{
We study the high-scale validity of two kinds of two Higgs doublet models (2HDM), namely, Type-II 
and Type-X,
but with a scalar SU(2) singlet dark matter (DM) candidate in addition in each case. The additional quartic couplings involving the DM particle in the scalar potential in both the scenarios bring in additional constraints from the requirement of perturbative unitarity and vacuum stability. DM relic density and direct search constraints play a crucial role in this analysis as the perturbative unitarity of the DM-Higgs portal couplings primarily decide the high scale validity of the model. We find that, within the parameter regions
thus restricted, the Type-II scenario must have a cut-off
at around  $10^6$ GeV, while the Type-X scenario admits of
validity upto the Planck scale. However, only those regions which are valid upto about $10^8$ GeV in Type-X
2HDM is amenable to detection at the High-luminosity LHC
(upto 3000 $fb^{-1}$), while most of the parameter space of the Type-II scenario mentioned above is likely to be detectable.

}

\preprint{DIAS-STP-23-13\\$\textrm{}$}

\DeclareUnicodeCharacter{2212}{-}
\begin{document}

\maketitle

\newpage

\section{Introduction}
\label{sec1}

The discovery and subsequent study of the 125-GeV scalar has almost decisively confirmed the spontaneous breakdown mechanism
in the standard electroweak model (SM). It is, however, still possible that more than one scalar SU(2) doublets participate in the electroweak symmetry breaking (EWSB) scheme. Two Higgs doublet models (2HDM)~\cite{Branco:2011iw} have thus become subjects of frequent investigation, motivated from various unanswered questions in the SM. There are various kinds of 2HDM, for each of which substantial regions of the parameter space are identified as consistent with observed phenomenology. This statement is particularly valid if one remains within the `alignment limit', much of which is accessible to accelerator searches.

Side by side, a question that constantly haunts physicists
is the origin of dark matter (DM) in our universe, which  strongly suggests physics beyond the SM (BSM), so long as DM is constituted of some yet unknown elementary particle(s). One possibility that is often explored in this context is whether a DM particle, especially a scalar one, can interact with the rest of the SM spectrum through the EWSB sector. Such `Higgs portal' scenarios, however, are strongly constrained from direct DM search data, because of the SM-like scalar contributions to the
spin-independent cross-section~\cite{Lopez-Honorez:2012tov,Greljo:2013wja,Fedderke:2014wda}. The restriction, however, become considerably relaxed, if the DM particle has appreciable interaction strength with a heavier neutral scalar in the 2HDM spectrum~\cite{Dey:2019lyr,Dey:2020tfq,Dey:2021alu}. In that case, not only does one have smaller, propagator-suppressed, contributions to direct search cross-sections, but it is also easier to reduce the tension between the Direct search limits~\cite{XENON:2018voc,LZ:2022ufs} and those from the relic density of the universe~\cite{Planck:2018vyg}, due to appropriate interface between more than one contributing channels and the larger number of parameters at one's disposal. The constraints as well as collider signatures of such {\it 2HDM + scalar DM} scenarios~\cite{Drozd:2014yla,Muhlleitner:2016mzt,Engeln:2020fld} have already been explored~\cite{Dey:2019lyr,Dey:2021alu}.

We go a step further in the current work. It is a natural question to ask as to what can be the ultraviolet (UV) behaviour of a  {\it 2HDM + scalar DM} scenario. Such a question is not only germane in the context of model-building but has also regarding implications in early universe issues, such as electroweak phase transition or the freeze-out of the scalar DM candidate. This issue becomes even more fascinating as the presence of DM turns out to crucially govern the high scale validity of the model.

Keeping such points in mind, we study the high-scale behaviour of such a theoretical scenario.
The running of the parameters via Renormalization Group Equations (RGE's) there lead to additional constraints arising from vacuum stability and perturbative unitarity at high scales. These bring in further restrictions of the allowed regions of the parameter space, over and above the ones already studied. It is thus important to know which parameter regions have chance of revealing themselves at the high-luminosity Large Hadron Collider (HL-LHC), corresponding to different upper limits of validity of this kind of a theory.
It is worth mentioning that the high-scale validity of Type-X 2HDM, especially of the parameter space giving rise to the observed anomalous magnetic moment of muon~\cite{Bennett:2006fi,Abi:2021gix,Albahri:2021ixb}, has been studied in a previous work~\cite{Dey:2021pyn}. It may be mentioned that the tension between theory and experiment in $g_{\mu}-2$ is claimed to have been relaxed on the basis of Lattice calculations~\cite{Borsanyi:2020mff,Ce:2022kxy,ExtendedTwistedMass:2022jpw,FermilabLatticeHPQCD:2023jof}. We nevertheless have taken a look at the Type-X 2HDM, since it still allows relatively light (pseudo)scalars, which has a bearing on the evolution of mass parameters upto high-scales. Furthermore, there have been studies on the vacuum stability as well as high scale validity of various other extended scalar sectors, like inert doublets and triplets in association with DM~\cite{Chakrabarty:2015yia,Ghosh:2017pxl,Bhattacharya:2019fgs}.

On the whole, the novel features of this study are as follows:

\begin{itemize}

\item We take up for our high-scale study two phenomenologically relevant 2HDM types, 
namely, Type-II (which occurs rather naturally in the supersymmetric SM)
and Type-X (which is of interest in the context of muon ($g-2$) ). Each of these scenarios are in addition augmented with one SU(2) singlet DM candidate, for which 
the scalar potential serves as the portal to SM physics.

\item For both these cases, the running of various quartic coupling strengths is studied.
A scan is made of the parameter space of each of the scenarios, and the allowed regions the parameter spaces are identified, considering in turn the constraints corresponding to different cut-off scales, and those coming from DM-related issues (mainly direct searches and relic density), in conjunction with the usual phenomenological limits. In particular, the parameter region in Type-X 2HDM
improving on the discrepancy in muon ($g-2$) is filtered out as an added requirement.

\item With the parameter regions thus narrowed down, the potential of capturing the signatures of such scenarios at the HL-LHC are commented upon.  

\end{itemize}

\noindent
The paper is organized as follows. We discuss the model and theoretical, experimental contraints as well as constraints from the dark matter sector on the model in Section~\ref{sec2}. In Section~\ref{sec3}, we discuss the RG running of all the couplings and demonstrate with a few benchmark points from Type-II and Type-X 2HDM. We identify the allowed parameters from the perspective of various high-scale validity and dark matter constraints, and discuss the interplay between the two in Section~\ref{sec4}. Finally, we summarize and conclude our discussion in Section~\ref{sec5}.

\section{Models and Constraints}
\label{sec2}

\subsection{Models}
As stated above, we concentrate on a two Higgs doublet model (2HDM), with an $SU(2)$ real singlet scalar dark matter candidate $S$. $S$ interacts with two higgs doublets $\Phi_{1,2}$. The scalar potential of the full scenario is

\begin{equation}
{\cal V} = {\cal V}_{2HDM} + \frac{1}{2} M_{S}^2 S^2 + \frac{\lambda_S}{4!} S^4 + \lambda_{S1} S^2 \Phi_1^{\dagger} \Phi_1 + \lambda_{S2} S^2 \Phi_2^{\dagger} \Phi_2. 
\end{equation}

where the terms in odd powers of $S$ are absent due to a $Z_2$ symmetry that stabilizes it. 

The most general scalar potential involving two scalar doublets in 2HDM is given as follows.

\begin{eqnarray} \label{V2HDM} {\cal V}_{2HDM} &=& m_{11}^2
(\Phi_1^{\dagger} \Phi_1) + m_{22}^2 (\Phi_2^{\dagger}
\Phi_2) - \left[m_{12}^2 (\Phi_1^{\dagger} \Phi_2 + \rm h.c.)\right]+ \frac{\lambda_1}{2}  (\Phi_1^{\dagger} \Phi_1)^2 +
\frac{\lambda_2}{2} (\Phi_2^{\dagger} \Phi_2)^2\nonumber \\
&&  + \lambda_3
(\Phi_1^{\dagger} \Phi_1)(\Phi_2^{\dagger} \Phi_2) + \lambda_4
(\Phi_1^{\dagger}
\Phi_2)(\Phi_2^{\dagger} \Phi_1)+ \left[\frac{\lambda_5}{2} (\Phi_1^{\dagger} \Phi_2)^2 + \rm
h.c.\right].
\end{eqnarray}

We assume CP-conservation, which is ensured by taking all $\lambda_i$'s and $m_{12}^2$ to be real.

The two complex Higgs doublets with hypercharge $Y = 1$ can be written as
\begin{equation}
\Phi_1=\left(\begin{array}{c} \phi_1^+ \\
\frac{1}{\sqrt{2}}\,(v_1+\phi_1^0+ia_1)
\end{array}\right)\,, \ \ \
\Phi_2=\left(\begin{array}{c} \phi_2^+ \\
\frac{1}{\sqrt{2}}\,(v_2+\phi_2^0+ia_2)
\end{array}\right).
\end{equation}
Where $v_1$ and $v_2$ are the vacuum expectation values of the two doublets, with $v^2 = v^2_1 + v^2_2 = (246~\rm GeV)^2$ and $\tan\beta=v_2 /v_1$. After EWSB, we obtain five physical states, two neutral CP-even scalars, the lighter of which will be called $h$,  and the heavier $H$, one neutral pseudoscalar $A$, and a pair of charged scalars $H^{\pm}$.


The above potential prevents mixing between $S$ and the scalar doublets as also any vacuum expectation value (VEV) for $S$. 
The mass of the DM candidate $S$ is given by, ${M^{phy}_{S}}^2 = M_{S}^2 + (\lambda_{S1}v_1^2+\lambda_{S2} v_2^2)$.

In order to suppress tree-level Flavour changing neutral current (FCNC), one needs to impose further a ${\cal {Z}}_2$ symmetry in the Yukawa sector. Depending on its nature there are four major Types of 2HDM's. Here we concentrate on Type-II and Type-X 2HDM.
In Type-II 2HDM, up-type quarks couple to one doublet, and down-type quarks and charged
leptons to the other doublet. Under this assumption, ${\cal L}_{Yukawa}$ becomes
\bea
\label{type2}
- {\cal L}_{Yukawa} &=&Y_{u2}\,\overline{Q}_L \, \tilde{{ \Phi}}_2 \,u_R
+\,Y_{d1}\,
\overline{Q}_L\,{\Phi}_1 \, d_R\, + \, Y_{\ell 1}\,\overline{L}_L \, {\Phi}_1\,e_R+\, \mbox{h.c.}\, \eea

\noindent
In Type-X 2HDM, on the other hand, the Yukawa interactions are given as
 \bea
\label{typex}
- {\cal L}_{Yukawa} &=&Y_{u2}\,\overline{Q}_L \, \tilde{{ \Phi}}_2 \,u_R
+\,Y_{d2}\,
\overline{Q}_L\,{\Phi}_2 \, d_R\, + \, Y_{\ell 1}\,\overline{L}_L \, {\Phi}_1\,e_R+\, \mbox{h.c.}\, \eea 

where $\Phi_1$ couples to leptons only and $\Phi_2$, only to quarks.
In Equation~\ref{type2} and \ref{typex},

$Q_L^T=(u_L\,,d_L)$, $L_L^T=(\nu_L\,,l_L)$, and
$\widetilde\Phi_{1,2}=i\tau_2 \Phi_{1,2}^*$. $Y_{u2}$,
$Y_{d1}$,$Y_{d2}$ and $Y_{\ell 1}$ are the couplings of the up, down quarks and leptons with the two doublets where family indices are suppressed.



It should also be noted that ${\cal {Z}}_2$ symmetry of Yukawa sector is present in the scalar potential as well, excepting for the soft-breaking term $m_{12}^2$.

\subsection{Theoretical constraints}
Theoretical constraints on the model include  perturbativity, unitarity and vacuum stability, all the way upto the energy scale which marks the upper limit of validity of the model. Effects of these constraints on various 2HDM parameter spaces have been studied in detail earlier~\cite{Bernon:2015qea,Bernon:2015wef,Crivellin:2013wna,Arbey:2017gmh,Hussain:2017tdf}.
it has been pointed out that large separation between $m_A$ and $m_{H^{\pm}}$ is disfavored from the requirement of vacuum stability and perturbativity.

\medskip

\noindent
$\bullet$ {\bf Vacuum stability:} We would like to check the boundedness from below condition of the scalar potential, which implies there exists no direction in the field space in which $\cal V \rightarrow -\infty$. This leads to the following conditions on the quartic couplings of the potential
~\cite{Deshpande:1977rw,Nie:1998yn,Gunion:2002zf}.
\begin{eqnarray}
\label{vs1}
\lambda_{1,2} > 0 \,, \\
\label{vs2}
\lambda_{S} > 0 \,, \\
\label{vsd}
\lambda_3  > -\sqrt{\lambda_1 \lambda_2} \,,\\
\label{vs3}
|\lambda_5| < \lambda_3 + \lambda_4 + \sqrt{\lambda_1 \lambda_2}\,\label{stab},\\
\lambda_{S1} > -\sqrt{\frac{1}{12} \lambda_S \lambda_1}, \lambda_{S2} > -\sqrt{\frac{1}{12} \lambda_S \lambda_2}. 
\label{stability}
\end{eqnarray} 
For negative $\lambda_{S1}$ or $\lambda_{S2}$ one additionally has to satisfy,

\begin{eqnarray}
\left(\frac{1}{12}\lambda_S \lambda_1 - \lambda_{S1}^2 \right) > 0, \\
\left(\frac{1}{12}\lambda_S \lambda_2 - \lambda_{S2}^2 \right) > 0, \\
- 2\lambda_{S1} \lambda_{S2} + \frac{1}{6} \lambda_S \lambda_3 > -\sqrt{4\left(\frac{1}{12}\lambda_S \lambda_1 - \lambda_{S1}^2 \right)\left(\frac{1}{12}\lambda_S \lambda_2 - \lambda_{S2}^2 \right)}, \\
- 2\lambda_{S1} \lambda_{S2} + \frac{1}{6} \lambda_S \left(\lambda_3 + \lambda_4 - |\lambda_5| \right) > -\sqrt{4\left(\frac{1}{12}\lambda_S \lambda_1 - \lambda_{S1}^2 \right)\left(\frac{1}{12}\lambda_S \lambda_2 - \lambda_{S2}^2 \right)}.
\label{stability_ls}
\end{eqnarray}

\noindent
$\bullet$ {\bf Perturbativity:} If 2HDM is a perturbative quantum field theory at a given scale, it would imply, all quartic couplings, involving the scalar mass eigenstates $C_{H_iH_jH_kH_l} < 4\pi$ and all Yukawa couplings $Y_j < \sqrt{4\pi}$. Further, unitarity bound on the tree level scattering
amplitude of the scalars and longitudinal parts of EW gauge bosons put an upper bound on the eigenvalues $|a_i|\leq 8\pi$ of the $2 \rightarrow 2$ scattering matrices~\cite{Arhrib:2000is,Kanemura:1993hm,Dicus:1992vj,Lee:1977yc,Lee:1977eg,Ginzburg:2005dt}.

The physical masses of the additional scalars can be expressed as: 

\begin{eqnarray}
m_A^2 &=& \frac{m_{12}^2}{\sin\beta \cos\beta} - \lambda_5 v^2,  \\
m_{H^{\pm}}^2 &\approx& m_A^2 + \frac{1}{2} v^2 (\lambda_5 - \lambda_4).
\label{massdiff}
\end{eqnarray}
\noindent
It is clear from Equation~\ref{massdiff} that $m_{H^{\pm}}^2 - m_A^2$ is proportional to $\lambda_5 - \lambda_4$ which should be less than $ \lambda_3 + \sqrt{\lambda_1 \lambda_2}$ from the requirement of boundedness from below (Equation~\ref{stab}). 
Therefore these conditions along with the requirement of perturbativity ie. $C_{H_iH_jH_kH_l} < 4\pi$ puts an upper limit on the mass square difference  $m_{H^{\pm}}^2 - m_A^2$.

The aforementioned constraints can be easily translated into those of the parameter space by expressing the quartic couplings into parameters of the physical basis i.e. masses of the scalars and the mixing angles as follows.

\begin{eqnarray}
\nonumber && \lambda_1 = \frac{m_H^2\cos^2\alpha+m_h^2\sin^2\alpha-m_{12}^2\tan\beta}{v^2\cos^2\beta},\\
\nonumber && \lambda_2 = \frac{m_H^2\sin^2\alpha+m_h^2\cos^2\alpha-m_{12}^2\cot\beta}{v^2\sin^2\beta},\\
\nonumber && \lambda_3 = \frac{(m_H^2-m_h^2)\cos\alpha \sin\alpha+2m_{H^\pm}^2\sin\beta \cos\beta-m_{12}^2}{v^2\sin\beta \cos\beta},\\
\nonumber && \lambda_4 = \frac{(m_A^2-2m_{H^\pm}^2)\sin\beta \cos\beta+m_{12}^2}{v^2\sin\beta \cos\beta},\\
&& \lambda_5 = \frac{m_{12}^2 - m_A^2\sin\beta \cos\beta}{v^2\sin\beta \cos\beta}.
\label{eq:paratran}
\end{eqnarray}
\noindent
One should note from the expression of $\lambda_1$ in Equation~\ref{eq:paratran} that, to have it in the perturbative limit, the soft ${\cal Z}_2$ breaking parameter $m_{12}^2 \approx \frac{m_H^2}{\tan \beta}$, especially when $m_H >> m_h$.

\medskip
\noindent

\subsection{Experimental constraints}
Now we briefly discuss experimental constraints on the model parameters. 
\medskip

\noindent
$\bullet$ {\bf Electroweak Precision measurements:} Electroweak precision measurements~\cite{BAAK:2014gga}, especially from the oblique parameters ($S,T,U)$~\cite{Lavoura:1993nq,Baak:2012kk}, put significant  constraint on 2HDM parameter space when considered at one-loop level, because of the presence of additional scalars.  In various earlier works, it is pointed out that the heavier neutral scalar ($H$) and charged scalar ($H^{\pm}$) masses should be closer to each other ($\Delta m \lsim 50$GeV), in order to avoid the breaking of custodial SU(2) symmetry~\cite{Broggio:2014mna,Haller:2018nnx,He:2001tp,Grimus:2007if,Bhattacharyya:2015nca}, and at the same time keep the pseudo-scalar mass less constrained. This limit on $\Delta m$ becomes stronger when $H$ and $H^{\pm}$become heavier ($\gsim 600$GeV).

\medskip

\noindent
$\bullet$ {\bf Collider bounds:} The CMS and ATLAS data from runs I and II, on the observed SM-like 125-GeV scalar provide measurements of its signal strengths in different channels with increasing precision~\cite{ATLAS:2016neq,CMS:2018uag,ATLAS:2019nkf}. The data shows significant agreement with the SM predictions of couplings and pushes the limit towards the so-called alignment limit, i.e, $(\beta-\alpha) \approx \frac{\pi}{2}$. 

The direct search of non-standard scalars also put severe constraints on the parameter space. For Type-X 2HDM, the $\tau\tau$ final state restricts our parameter space strongly~\cite{Khachatryan:2015tra,CMS:2016ncz}, mostly because we choose to work in the low scalar mass and large $\tan\beta$ region, owing its connection to $g_{\mu}-2$. For Type-II, on the other hand, the major constraint comes from $hh$ final state~\cite{CMS:2016jvt,CMS:2017aza}. LEP-limits on charged Higgs mass ($\gsim 80$GeV)~\cite{Abbiendi:2013hk} is imposed for both types. Type-II, in addition, gets constrained from B-physics observables which put a strong lower bound on the charged Higgs mass $m_{H^{\pm}} \gsim 600$ GeV~\cite{Crivellin:2013wna,Arbey:2017gmh,Hussain:2017tdf,Crivellin:2019dun}. All our chosen benchmarks in this work are consistent with the results of {\tt HiggsTools}, in particular, {\tt HiggsSignals}~\cite{Bechtle:2013xfa,Stal:2013hwa,Bechtle:2014ewa,Bechtle:2020uwn,Bahl:2022igd} and {\tt HiggsBounds}~\cite{Bechtle:2008jh,Bechtle:2011sb,Bechtle:2012lvg,Bechtle:2013wla,Bechtle:2015pma,Bechtle:2020pkv,Bahl:2021yhk}.

\subsection{Dark matter constraints} 

The WIMP-DM candidate of our model should satisfy the following constraints:\\
\medskip
$\bullet$ The thermal relic density should be consistent with the latest Planck data~\cite{Planck:2018vyg}. \\
$\bullet$ The DM-nucleon cross-section must be below 
the upper bound given by the latest LZ experiment~\cite{LZ:2022lsv}.\\
$\bullet$ Indirect detection constraints of Fermi-LAT experiments coming from isotropic gamma-ray data
and the gamma ray observations from dwarf spheroidal galaxies~\cite{Fermi-LAT:2015att} 
should be satisfied.\\

\section{Demonstration of running couplings with some benchmarks}
\label{sec3}
After discussing all the theoretical and experimental constraints, we examine high-scale validity of such models ({\it 2HDM + scalar}). First we list the RGE's for all gauge, Yukawa and scalar quartic couplings for our model at one-loop level. 
Though in rest of our work we used two-loop RGE's, we present the Equations for the one-loop RGE's to have an intuitive grasp on the key features of the running of relevant couplings. However, use has been made of the two-loop RGE's~\cite{Chowdhury:2015yja} only in the subsequent numerical analyses and the results that follow. The two-loop RGE's are shown in detail in Appendix~\ref{2looprge}. We have implemented the model and generated the one and two-loop RG equations in {\tt SARAH}~\cite{Staub:2009bi,Staub:2010jh,Staub:2012pb,Staub:2015kfa}. Subsequently, we evolved the couplings using the aforementioned RG equations using {\tt 2HDME}~\cite{Oredsson:2018vio}.

\subsection{The one-loop RGE's}
We begin by introducing the one-loop RGE's for the gauge couplings. Equation~\ref{gauge_rge} demonstrates that they constitute a stand-alone set at one-loop and, as a result, are the same for different types of 2HDM. We would like to mentions that GUT normalisation was not used while writing Equation~\ref{gauge_rge}.

\begin{align}
16 \pi ^2 \beta _{g_1} =&7 g_1^3, \nonumber \\
16 \pi ^2 \beta _{g_2} =&-3 g_2^3, \nonumber \\
16 \pi ^2 \beta _{g_3} =&-7 g_3^3.
\label{gauge_rge}
\end{align}

\noindent
However, the running of Yukawa couplings as well as quartic couplings pertaining to the scalar sector receive different contribution for different types of 2HDM's since the quark and lepton couplings with the scalar doublets play crucial role in these cases at one-loop level. We present the runnings of the aforementioned couplings for Type-II and Type-X 2HDM's.

\subsubsection{Type-II 2HDM}
We first concentrate on the RGE of Type-II 2HDM Yukawa couplings. 
Here, the superscripts $g$ and $Y$, stand for contributions from the gauge and Yukawa sector, respectively, to the running of the Yukawa couplings (taken here as real).

\begin{align}
16 \pi ^2 \beta _{Y_t}^{g} &=-\left( \frac{17}{12} g_1^2 +\frac{9}{4} g_2^2 +8 g_3^2\right) Y_t, \nonumber \\
16 \pi ^2 \beta _{Y_t}^{Y} &=\left( \frac{3}{2} Y_b^2 +\frac{9}{2} Y_t^2 + Y_{\tau}^2\right) Y_t -\left( Y_b^2 + Y_{\tau}^2\right) Y_t, \nonumber\\ 
16 \pi ^2 \beta _{Y_b}^{g} &=-\left( \frac{5}{12} g_1^2 +\frac{9}{4} g_2^2 +8 g_3^2\right) Y_b, \nonumber \\
16 \pi ^2 \beta _{Y_b}^{Y} &=\left( \frac{9}{2} Y_b^2 +\frac{3}{2} Y_t^2 + Y_{\tau}^2\right) Y_b - Y_t^2 Y_b, \nonumber\\
16 \pi ^2 \beta _{Y_\tau}^{g} &=-\left( \frac{15}{4}g_1^2 +\frac{9}{4}g_2^2\right) Y_\tau, \nonumber \\
16 \pi ^2 \beta _{Y_\tau}^{Y} &=\left(\frac{5}{2}Y_\tau^2 + 3 Y_b^2 \right) Y_\tau.
\label{yuk_rge}
\end{align}

\noindent
The resulting beta-function will be the sum of the gauge and Yukawa components. 

\begin{equation}
\beta _{Y}= \beta _{Y}^{g} + \beta _{Y}^{Y}.
\end{equation}
The relevant equations for the running of quartic couplings are given below. Here, the superscripts $b$ and $Y$ denote, respectively, bosonic(gauge couplings and quartic couplings) and Yukawa interactions, contributing to the running of $\lambda'$s.

\begin{align}
16 \pi ^2 \beta _{\lambda_1}^{b} =&\frac{3}{4} g_1^4 +\frac{3}{2} g_1^2 g_2^2 +\frac{9}{4} g_2^4 -3g_1^2 \lambda_1 -9 g_2^2 \lambda_1 +12 \lambda_1^2 +4 \lambda_3^2 +4 \lambda_3 \lambda_4 +2\lambda_4^2 +2\lambda_5^2 +  4 \lambda_{S1}^2, \nonumber \\
16 \pi ^2 \beta _{\lambda_1}^{Y} =& -4 Y_\tau^4 +4 Y_\tau^2 \lambda_1 -12 Y_b^4 + 12 Y_b^2 \lambda_1, \nonumber\\
16 \pi ^2 \beta _{\lambda_2}^{b} =& \frac{3}{4} g_1^4 +\frac{3}{2} g_1^2 g_2^2 +\frac{9}{4} g_2^4 -3g_1^2 \lambda_2 -9 g_2^2 \lambda_2 +12 \lambda_2^2 +4 \lambda_3^2 +4 \lambda_3 \lambda_4 +2\lambda_4^2 +2\lambda_5^2 + 4 \lambda_{S2}^2, \nonumber\\
16 \pi ^2 \beta _{\lambda_2}^{Y} =& -12 Y_t^4 + 12 Y_t^2 \lambda_2, \nonumber\\
16 \pi ^2 \beta _{\lambda_3}^{b} =& \frac{3}{4} g_1^4 -\frac{3}{2} g_1^2 g_2^2 +\frac{9}{4} g_2^4 -3g_1^2 \lambda_3 -9 g_2^2 \lambda_3 +(\lambda_1 +\lambda_2 ) \left( 6  \lambda_3 +2\lambda_4\right) +4 \lambda_3^2 +2\lambda_4^2 +2\lambda_5^2 +  4 \lambda_{S1} \lambda_{S2}, \nonumber\\
16 \pi ^2 \beta _{\lambda_3}^{Y} =& \left(6 Y_b^2 +6 Y_t^2 +2 Y_\tau^2\right) \lambda_3 - 12 Y_b^2 Y_t^2, \nonumber\\
16 \pi ^2 \beta _{\lambda_4}^{b} =& 3 g_1^2 g_2^2 -\left( 3 g_1^2 +9 g_2^2 \right) \lambda_4 +2 \lambda_1 \lambda_4+2 \lambda_2 \lambda_4+8 \lambda_3 \lambda_4+4 \lambda_4^2+8 \lambda_5^2, \nonumber\\
16 \pi ^2 \beta _{\lambda_4}^{Y} =& \left( 6 Y_b^2 +6 Y_t^2 +2 Y_\tau^2\right) \lambda_4 + 12 Y_b^2 Y_t^2, \nonumber\\
16 \pi ^2 \beta _{\lambda_5}^{b} =& \left(-3 g_1^2 -9 g_2^2 +2\lambda_1 +2\lambda_2 +8 \lambda_3 +12 \lambda_4\right) \lambda_5, \nonumber\\
16 \pi ^2 \beta _{\lambda_5}^{Y} = & \left(6 Y_b^2 +6 Y_t^2 +2Y_\tau^2\right) \lambda_5, \nonumber\\
 16 \pi ^2 \beta _{\lambda_S}^{b} = & 3 \left(16 \lambda_{S1}^2 + 16 \lambda_{S2}^2 + \lambda_S^2\right), \nonumber\\
 16 \pi ^2 \beta _{\lambda_S}^{Y} = & 0, \nonumber\\
16 \pi ^2 \beta _{\lambda_{S1}}^{b} = & -\frac{3}{2} g_1^2 \lambda_{S1} - \frac{9}{2} g_2^2 \lambda_{S1} + 6 \lambda_1 \lambda_{S1} + \lambda_S \lambda_{S1} + 4 \lambda_3 \lambda_{S2} + 2 \lambda_4  \lambda_{S2} + 8 \lambda_{S1}^2, \nonumber\\
16 \pi ^2 \beta _{\lambda_{S1}}^{Y} =  &  6 \lambda_{S1} Y_b^2 + 2 \lambda_{S1} Y_{\tau}^2,
\nonumber\\
16 \pi ^2 \beta _{\lambda_{S2}}^{b} = & -\frac{3}{2} g_1^2 \lambda_{S2} - \frac{9}{2} g_2^2 \lambda_{S2} + 6 \lambda_2 \lambda_{S2} + \lambda_S \lambda_{S2} + 4 \lambda_3 \lambda_{S1} + 2 \lambda_4  \lambda_{S1} + 8 \lambda_{S2}^2, \nonumber\\
16 \pi ^2 \beta _{\lambda_{S2}}^{Y} =  & 6 \lambda_{S2} Y_t^2.
\label{lambda_rge_type2}
\end{align}

\noindent
Like before, the actual beta-function will be the sum of the bosonic and Yukawa components. 

\begin{equation}
\beta _{\lambda}= \beta _{\lambda}^{b} + \beta _{\lambda}^{Y}.
\end{equation}

\noindent
Comparing with Equation~\ref{type2}, we would like to make the following identifications for type-II 2HDM.

$~~~~~~~~~~~~~~~~~~~~~~~~~~~~~~~Y_{u2} = Y_t,~~~ Y_{d1} = Y_b~~$and$~~Y_{\ell 1} = Y_{\tau}$.

\subsubsection{Type-X 2HDM}
Next we focus on the RGE of the Yukawa couplings in Type-X 2HDM. 
Here too, the superscripts $g$ and $Y$ stand for contributions from gauge and Yukawa sectors, respectively.

\begin{align}
16 \pi ^2 \beta _{Y_t}^{g} &=-\left( \frac{17}{12} g_1^2 +\frac{9}{4} g_2^2 +8 g_3^2\right) Y_t, \nonumber \\
16 \pi ^2 \beta _{Y_t}^{Y} &=\left( \frac{3}{2} Y_b^2 +\frac{9}{2} Y_t^2 \right) Y_t, \nonumber\\ 
16 \pi ^2 \beta _{Y_b}^{g} &=-\left( \frac{5}{12} g_1^2 +\frac{9}{4} g_2^2 +8 g_3^2\right) Y_b, \nonumber \\
16 \pi ^2 \beta _{Y_b}^{Y} &=\left( \frac{9}{2} Y_b^2 +\frac{3}{2} Y_t^2 \right) Y_b, \nonumber\\
16 \pi ^2 \beta _{Y_\tau}^{g} &=-\left( \frac{15}{4}g_1^2 +\frac{9}{4}g_2^2\right) Y_\tau, \nonumber \\
16 \pi ^2 \beta _{Y_\tau}^{Y} &=\frac{5}{2}Y_\tau^3.
\label{yuk_rge}
\end{align}

\noindent
The gauge and Yukawa components will be added to provide the final beta-function.  

\begin{equation}
\beta _{Y}= \beta _{Y}^{g} + \beta _{Y}^{Y}.
\end{equation}

\noindent
We present next the running of scalar quartic couplings below. The superscripts $b$ and $Y$ bear similar meaning as in the type-II case.

\begin{align}
16 \pi ^2 \beta _{\lambda_1}^{b} =&\frac{3}{4} g_1^4 +\frac{3}{2} g_1^2 g_2^2 +\frac{9}{4} g_2^4 -3g_1^2 \lambda_1 -9 g_2^2 \lambda_1 +12 \lambda_1^2 +4 \lambda_3^2 +4 \lambda_3 \lambda_4 +2\lambda_4^2 +2\lambda_5^2   + 4\lambda_{S1}^2, \nonumber \\
16 \pi ^2 \beta _{\lambda_1}^{Y} =& -4 Y_\tau^4 +4 Y_\tau^2 \lambda_1, \nonumber\\
16 \pi ^2 \beta _{\lambda_2}^{b} =& \frac{3}{4} g_1^4 +\frac{3}{2} g_1^2 g_2^2 +\frac{9}{4} g_2^4 -3g_1^2 \lambda_2 -9 g_2^2 \lambda_2 +12 \lambda_2^2 +4 \lambda_3^2 +4 \lambda_3 \lambda_4 +2\lambda_4^2 +2\lambda_5^2  + 4\lambda_{S2}^2, \nonumber\\
16 \pi ^2 \beta _{\lambda_2}^{Y} =& -12 Y_b^4 -12 Y_t^4 +\left( 12 Y_b^2 +12 Y_t^2 \right) \lambda_2, \nonumber\\
16 \pi ^2 \beta _{\lambda_3}^{b} =& \frac{3}{4} g_1^4 -\frac{3}{2} g_1^2 g_2^2 +\frac{9}{4} g_2^4 -3g_1^2 \lambda_3 -9 g_2^2 \lambda_3 + (\lambda_1 +\lambda_2 ) \left( 6  \lambda_3 +2\lambda_4\right) +4 \lambda_3^2 +2\lambda_4^2 +2\lambda_5^2  + 4\lambda_{S1} \lambda_{S2}, \nonumber\\
16 \pi ^2 \beta _{\lambda_3}^{Y} =& \left(6 Y_b^2 +6 Y_t^2 +2 Y_\tau^2\right) \lambda_3, \nonumber\\
16 \pi ^2 \beta _{\lambda_4}^{b} =& 3 g_1^2 g_2^2 -\left( 3 g_1^2 +9 g_2^2 \right) \lambda_4 +2 \lambda_1 \lambda_4+2 \lambda_2 \lambda_4+8 \lambda_3 \lambda_4+4 \lambda_4^2+8 \lambda_5^2, \nonumber\\
16 \pi ^2 \beta _{\lambda_4}^{Y} =& \left( 6 Y_b^2 +6 Y_t^2 +2 Y_\tau^2\right) \lambda_4, \nonumber\\
16 \pi ^2 \beta _{\lambda_5}^{b} =& \left(-3 g_1^2 -9 g_2^2 +2\lambda_1 +2\lambda_2 +8 \lambda_3 +12 \lambda_4\right) \lambda_5, \nonumber\\
16 \pi ^2 \beta _{\lambda_5}^{Y} = & \left(6 Y_b^2 +6 Y_t^2 +2Y_\tau^2\right) \lambda_5, \nonumber\\
16 \pi ^2 \beta _{\lambda_S}^{b} = &  3 \left(16 \lambda_{S1}^2 + 16 \lambda_{S2}^2 + \lambda_S^2\right), \nonumber\\
16 \pi ^2 \beta _{\lambda_S}^{Y} = & 0, \nonumber\\
16 \pi ^2 \beta _{\lambda_{S1}}^{b} = & -\frac{3}{2} g_1^2 \lambda_{S1} - \frac{9}{2} g_2^2 \lambda_{S1} + 6 \lambda_1 \lambda_{S1} + \lambda_S \lambda_{S1} + 4 \lambda_3 \lambda_{S2} + 2 \lambda_4  \lambda_{S2} + 8 \lambda_{S1}^2, \nonumber\\
16 \pi ^2 \beta _{\lambda_{S1}}^{Y} =  &  2 \lambda_{S1} Y_{\tau}^2,
\nonumber\\
16 \pi ^2 \beta _{\lambda_{S2}}^{b} = &  -\frac{3}{2} g_1^2 \lambda_{S2} - \frac{9}{2} g_2^2 \lambda_{S2} + 6 \lambda_2 \lambda_{S2} + \lambda_S \lambda_{S2} + 4 \lambda_3 \lambda_{S1} + 2 \lambda_4  \lambda_{S1} + 8 \lambda_{S2}^2, \nonumber\\
16 \pi ^2 \beta _{\lambda_{S2}}^{Y} =  &  6 \lambda_{S2} Y_t^2 + 6 \lambda_{S2} Y_b^2.
\label{lambda_rge_typex}
\end{align}

\noindent
The actual beta-function, as before, will be the sum of the bosonic and Yukawa parts. 

\begin{equation}
\beta _{\lambda}= \beta _{\lambda}^{b} + \beta _{\lambda}^{Y}.
\end{equation}

\noindent
In case of Type-X 2HDM, comparing with Equation~\ref{typex}, we make the following identifications.

$~~~~~~~~~~~~~~~~~~~~~~~~~~~~~~~~~~Y_{u2} = Y_t,~~~ Y_{d2} = Y_b~~$and$~~Y_{\ell 1} = Y_{\tau}$.

\subsection{Choice of benchmarks and the running of quartic couplings}

In this subsection we will try to understand the pattern of the running of different quartic couplings for our case. The pattern of running for different 2HDM's are already studied extensively in the literature~\cite{Chowdhury:2015yja,Dey:2021pyn}. Here our main goal is to see how the scalar DM affects the cut-off scales $\Lambda^{Cut-off}_{UV}$, where vacuum stability or unitarity or perturbativity breaks. We would also like to compare this scenario with the 2HDM scenarios. We choose a few different benchmark points, four for each Type-II and Type-X with different values of $\lambda_S$, $\lambda_{S1}$ and $\lambda_{S2}$, presented in Table~\ref{type2_bp} and Table~\ref{typex_bp} respectively. We show the their running upto their respective cut-off scales  $\Lambda^{Cut-off}_{UV}$. Here we use two-loop RGE's of different quartic couplings presented in Appendix~\ref{2looprge}. Please also note that at this stage, the constraints from DM direct search or relic density have not been taken into account.  These will come  {\it post facto}, as exemplified in~Section~\ref{sec4}.

\subsubsection{Type-II 2HDM}
For Type-II 2HDM benchmark points we choose scalar masses and mixing angles which are allowed by all the theoretical and experimental constraints, as presented below.

\noindent
$m_h$ = 125.0 GeV, $m_H$ = 588.0 GeV, $m_A$ = 588 GeV, $m_H^{\pm}$ = 610 GeV, $m^2_{12}$ = 35776.44128 GeV$^2$\footnote{Some justification of the places after decimal retained in the various parameters is in order. The parameter $m_{12}^2$ decides rather sensitively the $m_H-m_h$ splitting and consequently, the scale at which perturbative unitarity breaks. We have noted that a shift in $m_{12}^2$ by a small amount in decimal places sometimes shifts the cut-off scale by about an order of magnitude. Keeping this in mind, we have retained the values of various parameters upto various decimal places which conform to the results yielded by {\tt 2HDME}.}, $\tan \beta$ = 9.6, $\sin(\beta-\alpha)$ = 0.998, which in the general basis leads to $\lambda_1$ = 1.60, $\lambda_2$ = 0.21, $\lambda_3$ = 3.15, $\lambda_4$ = 0.42, $\lambda_5$ = 0.03. Alongside, we choose four sets for DM sector parameters $\lambda_S$, $\lambda_{S1}$ and $\lambda_{S2}$(see Table ~\ref{type2_bp}). 

\begin{table}[!hptb]
\begin{center}
\begin{tabular}{|c|c|c|c|c|}
\hline
& Type-II BP1 & Type-II BP2 & Type-II BP3 & Type-II BP4\\
\hline
$\lambda_S$ & 0.0 & 0.1 & 1.0 $\times 10^{-6}$ & 0.1\\
\hline
$\lambda_{S1}$ & 0.0 & 0.3 & 3.0 $\times 10^{-6}$ & 2.5\\
\hline
$\lambda_{S2}$ & 0.0 & 2.0 & 1.5 & 2.5\\
\hline
$\Lambda^{Cut-off}_{UV}$ (2-loop) in GeV & 5.17 $\times 10^{3}$ & 4.27 $\times 10^{3}$ & 4.79 $\times 10^{3}$ & 1.39 $\times 10^{3}$\\
\hline
\end{tabular}
\caption{BP's for Type-II 2HDM with singlet scalar DM}
\label{type2_bp}
\end{center}
\end{table}

\noindent
BP1 corresponds to the usual Type-II 2HDM case, without the addition of scalar singlet DM. On the other hand, for BP2 our $\lambda_{S2}$ is larger than both $\lambda_{S1}$ and $\lambda_{S}$, which are kept at moderate values. In case of BP3, $\lambda_{S2}$ is taken to be much larger than $\lambda_S$ and $\lambda_{S1}$, both of which are kept at extremely small values. Finally in BP4, both of the $\lambda_{S1}$ and $\lambda_{S2}$ are larger compared to $\lambda_S$ and $\lambda_S$ is chosen at a moderate value.

\begin{figure}[!hptb]
\floatsetup[subfigure]{captionskip=10pt}
    \begin{subfigure}{.44\linewidth}
    \centering
    \includegraphics[width=10.0cm, height=7.0cm]{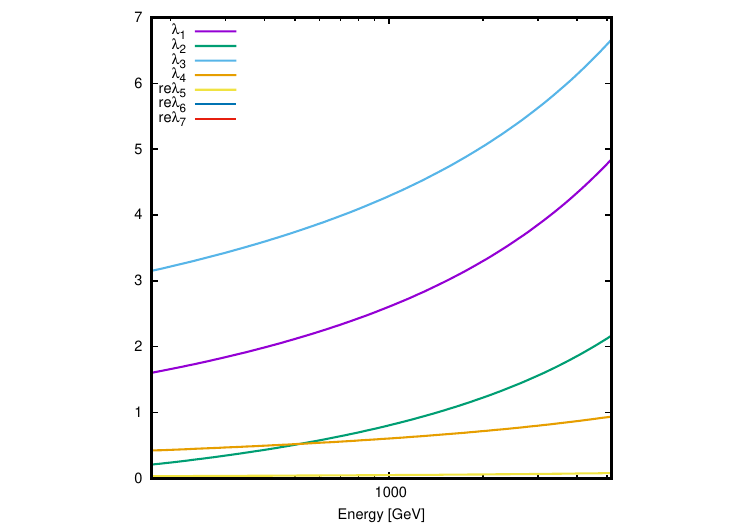}
    \caption{BP1:$\Lambda^{Cut-off}_{UV}$ = 5.175 TeV (Unitarity breakdown)}\label{fig:image11}
    \end{subfigure} %
    \qquad
    \begin{subfigure}{.44\linewidth}
    \centering
    \includegraphics[width=10.0cm, height=7.0cm]{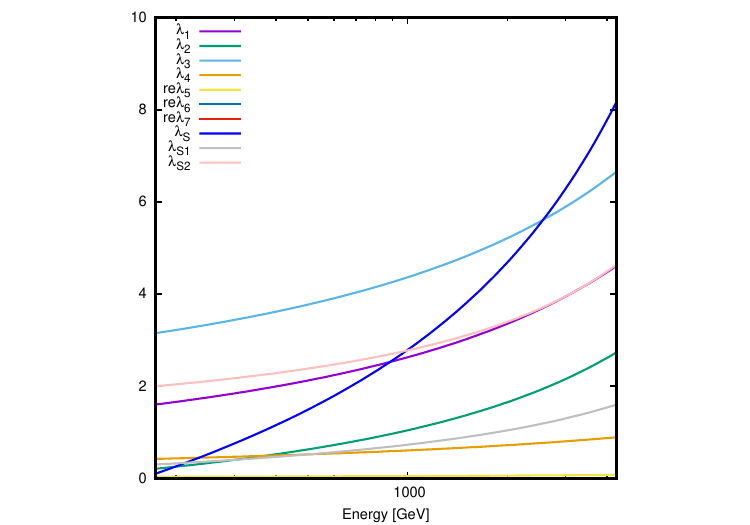}
    \caption{BP2:$\Lambda^{Cut-off}_{UV}$ = 4.266 TeV (Unitarity breakdown)}\label{fig:image12}
   \end{subfigure}
\\[2ex]
    \begin{subfigure}{.44\linewidth}
    \centering
    \includegraphics[width=10.0cm, height=7.0cm]{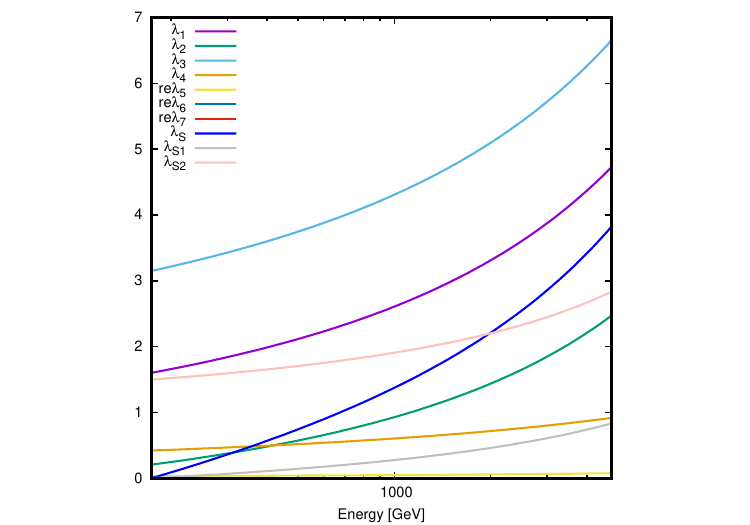}
    \caption{BP3:$\Lambda^{Cut-off}_{UV}$ = 4.790 TeV (Unitarity breakdown)}\label{fig:image13}
    \end{subfigure} %
    \qquad
    \begin{subfigure}{.44\linewidth}
    \centering
    \includegraphics[width=10.0cm, height=7.0cm]{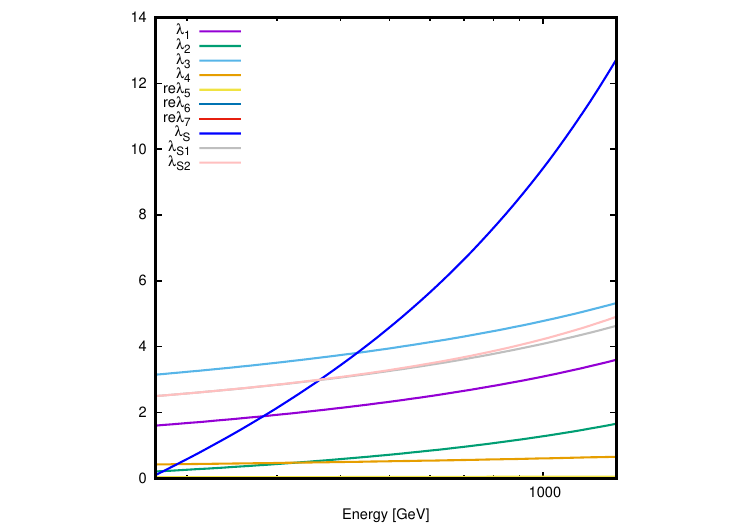}
    \caption{BP4:$\Lambda^{Cut-off}_{UV}$ = 1.393 TeV (Perturbativity breakdown)}\label{fig:image14}
   \end{subfigure}

\RawCaption{\caption{\it RG running of quartic couplings for benchmarks (a) BP1, (b) BP2, (c) BP3, (d) BP4 for Type-II with singlet scalar DM scenario. In all cases two-loop RGE's have been used.}
\label{bp123}}
\end{figure}

Figure~\ref{bp123} represents the two-loop RG running of various quartic couplings for Type-II scenario with starting scale set at top quark pole mass. In Figure~\ref{bp123}(a), $\lambda_S$, $\lambda_{S1}$ and $\lambda_{S2}$ is set to zero at EW scale. As the RGE's of these three $\lambda$'s are always proportional to one of these three $\lambda$, all of three remain zero at any energy scale even upto two-loop renormalization. Therefore, in this case the perturbative unitarity of the quartic couplings $\lambda_3$ determines the scale of validity. 

In case of BP2 and BP4, where $\lambda_S$ has moderate values, the cut-off scale is determined by the perturbativity of $\lambda_S$. On the other hand, for BP3, the cut-off scale $\Lambda^{Cut-off}_{UV}$ is determined by the perturbative unitarity of quartic coupling $\lambda_3$, since in this case $\lambda_S$ is extremely small.

In all the benchmarks in Figure~\ref{bp123}, the quartic couplings increase with energy. Since $\lambda_3$ is largest among all the quartic couplings at EW scale for our benchmark and its running involves the factor $4 \lambda_{S1} \lambda_{S2}$ (see Equation~\ref{lambda_rge_type2}), its perturbative unitarity breaks at much lower scale for for BP2, BP3 and BP4, compared to BP1 (normal 2HDM), as long as $\lambda_{S1}$ and $\lambda_{S2}$ are of the same sign. In such cases, normal 2HDM type-II scenario (BP1) naturally corresponds to largest cut-off scale (see Figure~\ref{bp123}(a)). If $\lambda_{S1}$ and $\lambda_{S2}$ are of different sign, the vacuum stability breaks down early on. Therefore even in that case, inclusion of DM worsens the high scale validity of normal 2HDM.

We would like to note that the benchmark points chosen here correspond to a large $\lambda_3$ at the EW scale, for which the cut-off scale of the model turns out to be in TeV scale (see Figure~\ref{bp123}). However, this is not a generic feature and the model can be valid to much higher scales, as we show in the scan in the next section.

Furthermore, $\lambda_S$ can also play an important role in the breaking of perturbative unitarity. The running of $\lambda_S$ is determined by $\lambda_S$, $\lambda_{S1}$ and $\lambda_{S2}$. It is therefore clear that for BP4 the perturbativity breaks down at a much lower scale compared to BP2, since the values of $\lambda_{S1}$ and $\lambda_{S2}$ for BP4 are largest among all benchmarks.


\subsubsection{Type-X 2HDM}

For Type-X 2HDM we choose the following benchmarks. Similar to Type-II case, here too, all the masses and mixing angles are allowed by theoretical and experimental constraints,
\noindent
$m_h$ = 93.6 GeV, $m_H$ = 125.0 GeV, $m_A$ = 15.8 GeV, $m_H^{\pm}$ = 135.0, $m^2_{12}$ = 393.28757 GeV$^2$\footnote{Here too, the retention of the number of places after decimal for various parameters is guided by the same consideration as that in the case of Type-II 2HDM.}, $\tan \beta$ = 22.0 $\sin(\beta-\alpha)$ = 0.006.
Our chosen masses and mixing angles in the physical basis leads to the following quartic couplings in the flavor basis,
$\lambda_1$ = 1.03, $\lambda_2$ = 0.26, $\lambda_3$ = 0.59, $\lambda_4$ = -0.45, $\lambda_5$ = 0.14.

One should note that unlike the Type-II case, here, our SM-like Higgs of 125 GeV mass is the second lightest CP-even scalar, which implies mixing angle $\sin(\beta-\alpha) << 1$,  This region is favored from the simultaneous requirement of high scale validity and the observed $g_{\mu}-2$~\cite{Dey:2021pyn}. The complementary region where 125 GeV Higgs is the lightest, has been studied in~\cite{Dey:2021pyn} without the inclusion of DM. Note that this is just for illustration; in Section~\ref{sec4}, we present a more general parameter scan. Furthermore, we choose four sets for DM sector parameters $\lambda_S$, $\lambda_{S1}$ and $\lambda_{S2}$ (see Table ~\ref{typex_bp}). 


\begin{table}[!hptb]
\begin{center}
\begin{tabular}{|c|c|c|c|c|}
\hline
& Type-X BP5 & Type-X BP6 & Type-X BP7 & Type-X BP8\\
\hline
$\lambda_S$ & 0.0 & 1.0 $\times 10^{-6}$ & 1.0 $\times 10^{-6}$ & 4.0\\
\hline
$\lambda_{S1}$ & 0.0 & 0.69 & 3.0 $\times 10^{-6}$ & -0.24\\
\hline
$\lambda_{S2}$ & 0.0 & 3.0 $\times 10^{-6}$ & 0.69 & 0.24\\
\hline
$\Lambda^{Cut-off}_{UV}$ (2-loop) in GeV & 2.27 $\times 10^{8}$ & 4.49 $\times 10^{7}$ & 2.02 $\times 10^{8}$ & 1.85 $\times 10^{4}$\\
\hline
\end{tabular}
\caption{BP's for Type-X 2HDM with singlet scalar DM}
\label{typex_bp}
\end{center}
\end{table}

\noindent
Here too, BP5 represents the usual Type-X 2HDM scenario where the values of new couplings pertaining to DM sector ($\lambda_S$, $\lambda_{S1}$ and $\lambda_{S2}$) are set to zero. For BP6, $\lambda_{S1}$ is chosen to be much larger than the other two couplings, where in BP7 $\lambda_{S2}$ is much larger than the other two. On the other hand, in BP8, both of the $\lambda_{S1}$ and $\lambda_{S2}$ are chosen to be moderate, but smaller compared to $\lambda_S$ and they have similar magnitude with opposite sign. 

\begin{figure}[!hptb]
\floatsetup[subfigure]{captionskip=10pt}
    \begin{subfigure}{.44\linewidth}
    \centering
    \includegraphics[width=10.0cm, height=7.0cm]{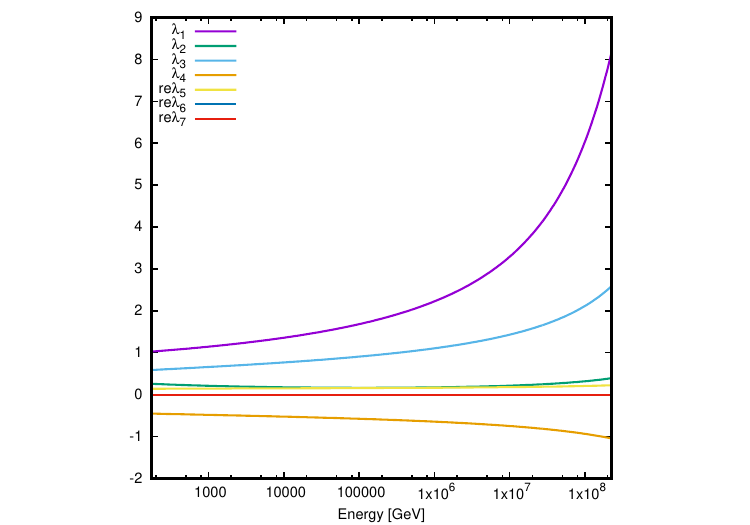}
    \caption{BP5 : $\Lambda^{Cut-off}_{UV}$ = 2.273$\times 10^5$ TeV (Unitarity breakdown)}\label{fig:image11x}
    \end{subfigure} %
    \qquad
    \begin{subfigure}{.44\linewidth}
    \centering
    \includegraphics[width=10.0cm, height=7.0cm]{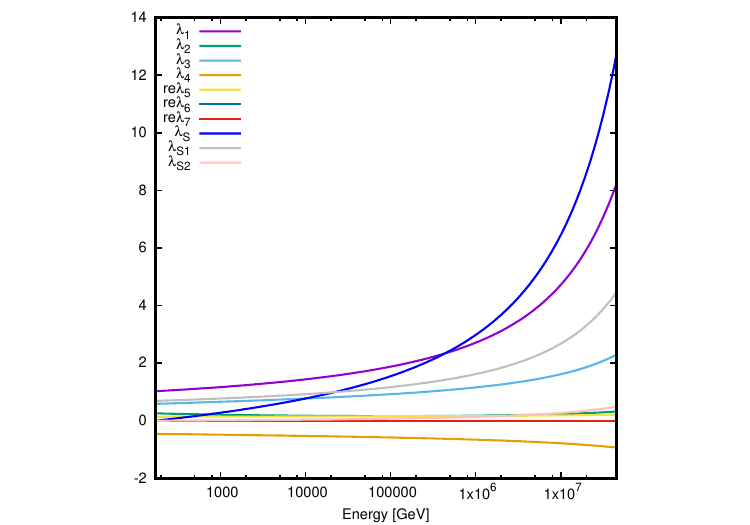}
    \caption{BP6 : $\Lambda^{Cut-off}_{UV}$ = 4.495$\times 10^4$ TeV (Perturbativity breakdown)}\label{fig:image12x}
   \end{subfigure}
\\[2ex]
    \begin{subfigure}{.44\linewidth}
    \centering
    \includegraphics[width=10.0cm, height=7.0cm]{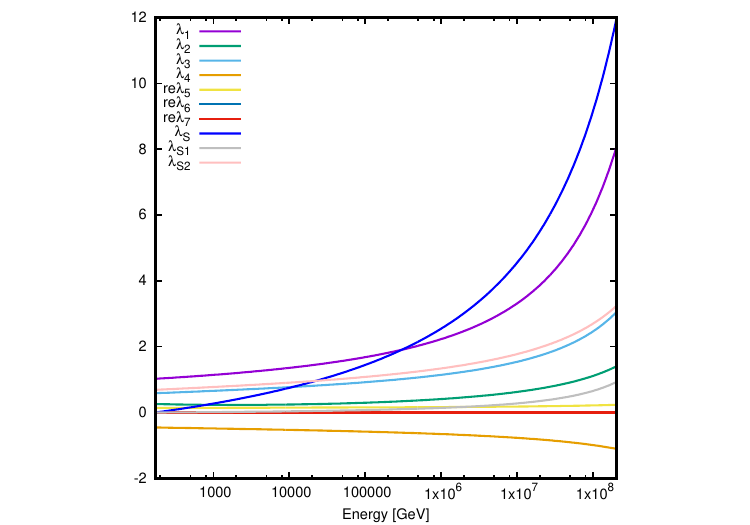}
    \caption{BP7 : $\Lambda^{Cut-off}_{UV}$ = 2.025$\times 10^5$ TeV (Unitarity breakdown)}\label{fig:image13x}
    \end{subfigure} %
    \qquad
    \begin{subfigure}{.44\linewidth}
    \centering
    \includegraphics[width=10.0cm, height=7.0cm]{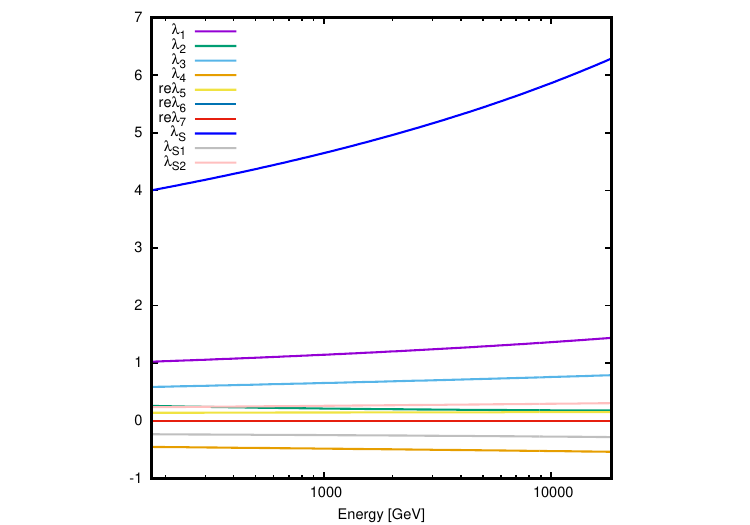}
    \caption{BP8 : $\Lambda^{Cut-off}_{UV}$ = 18.49 TeV (Stability breakdown)}\label{fig:image14x}
   \end{subfigure}

\RawCaption{\caption{\it RG running of quartic couplings for the benchmarks (a) BP5, (b) BP6, (c) BP7 and (d) BP8 for Type-X with singlet scalar DM scenario. In all cases two-loop RGE's have been used.}
\label{bp1231x}}
\end{figure}

\noindent
Figure~\ref{bp1231x} represents the one-loop RG evolution of various quartic couplings for Type-X scenario, with the initial scale set at top quark pole mass. In BP5 (Figure~\ref{bp1231x}(a)) $\lambda_S$, $\lambda_{S1}$ and $\lambda_{S2}$ are set to zero at EW scale and since the RGE's of this three couplings are proportional to at least one of these three couplings (same as Type-II scenario), they remain zero at any higher energy scale for BP5 at one-loop. Therefore, 
at one-loop level, for BP5 the cut-off scale $\Lambda^{Cut-off}_{UV}$ is determined by the perturbativity and unitarity of quartic couplings of Type-X 2HDM, especially $\lambda_1$, since in our chosen benchmark $\lambda_1$ is the largest.


In Figures~\ref{bp1231x}(a),~\ref{bp1231x}(b) and ~\ref{bp1231x}(c), all the quartic couplings increase with energy except $\lambda_4$, while in Figure~\ref{bp1231x}(d) both $\lambda_4$ and $\lambda_{S1}$ decrease with energy. $\lambda_1$ is largest among all at EW scale for first three benchmarks and its running majorly depends on the factor $4\lambda^2_{S1}$, when $\lambda^2_{S1}$ has a non-zero value at EW scale. The perturbative unitarity breaks at a lower scale for BP6 (Figure~\ref{bp1231x}(b)) compared BP7 (Figure~\ref{bp1231x}(c)), because of higher value of $\lambda_{S1}$ in case of BP6. Similar to Type-II, here too, normal 2HDM Type-X scenario, namely BP5 pertains to the highest cut-off scale. 
Interestingly, since the running of $\lambda_S$ is symmetric in $\lambda_{S1}$ and $\lambda_{S2}$, $\lambda_S$ runs identically for BP6 and BP7.

On the other hand, $\lambda_S$ increases much faster than other $\lambda$'s for BP8 at two-loop (Figure~\ref{bp1231x}(d)). BP8 shows a distinct behaviour because of the negative sign of $\lambda_{S1}$. In this case, the most stringent constraint comes  from vacuum stability, if one or both $\lambda_{S1}$ or $\lambda_{S2}$ are chosen negative, which breaks the stability at much smaller scale, which is the case with BP8. \\

In our earlier work~\cite{Dey:2021pyn}, we have discussed in detail the running of quartic couplings of Type-X 2HDM. It is not difficult to understand how the presence of the SM-DM and DM-DM couplings affect the allowed parameter space obtained there. We can see from Equations~\ref{lambda_rge_type2} and \ref{lambda_rge_typex} , $\lambda_1$, $\lambda_2$, always gets positive contribution, in terms of $4\lambda_{S1}^2$ and $4\lambda_{S2}^2$ respectively, while $\lambda_3$ receives positive or negative contribution ($4\lambda_{S1}\lambda_{S2}$), depending on relative size of $\lambda_{S1}$ and $\lambda_{S2}$. Therefore, if $\lambda_3$, $\lambda_{S1}$ and $\lambda_{S2}$ are considerably large compared to all other quartic couplings and $\lambda_{S1}$ and $\lambda_{S2}$ come with a relative negative sign, there is a possibility of getting a more relaxed parameter space in terms of perturbative unitarity. But in that case, the vacuum stability will be at stake (similar to BP8) and the final allowed parameter space will be more restricted compared to 2HDM parameter space. 


It is quite apparent that a WIMP like scalar DM, having sizable portal coupling, restricts the high scale validity of the two Higgs doublet models significantly, while interestingly, a FIMP (Feebly Interacting Massive Particle) like scalar singlet having tiny portal-couplings won't affect the high scale validity of the model so much. 

\section{Study of model parameter space}

\label{sec4}

\subsection{Regions of high-scale validity}

After discussing the RG evolutions of all the relevant
couplings in the model, we proceed to scan the model
parameter space and look for points that satisfy all the
theoretical constraints, namely perturbativity, unitarity, and
vacuum stability up to cutoff scale $\Lambda_{UV}^{cut-off}$. We have chosen four different scales in this context, namely, $10^4$,$10^8$,$10^{16}$ and $10^{19}$ GeV, and present the corresponding allowed regions in the parameter space, spanned by the DM-sector couplings $\lambda_{S1},\lambda_{S2}$ and $\lambda_S$.

\begin{figure}[!hptb]
\floatsetup[subfigure]{captionskip=10pt}
    \begin{subfigure}{.44\linewidth}
    \centering
    \includegraphics[width=8.0cm, height=6.0cm]{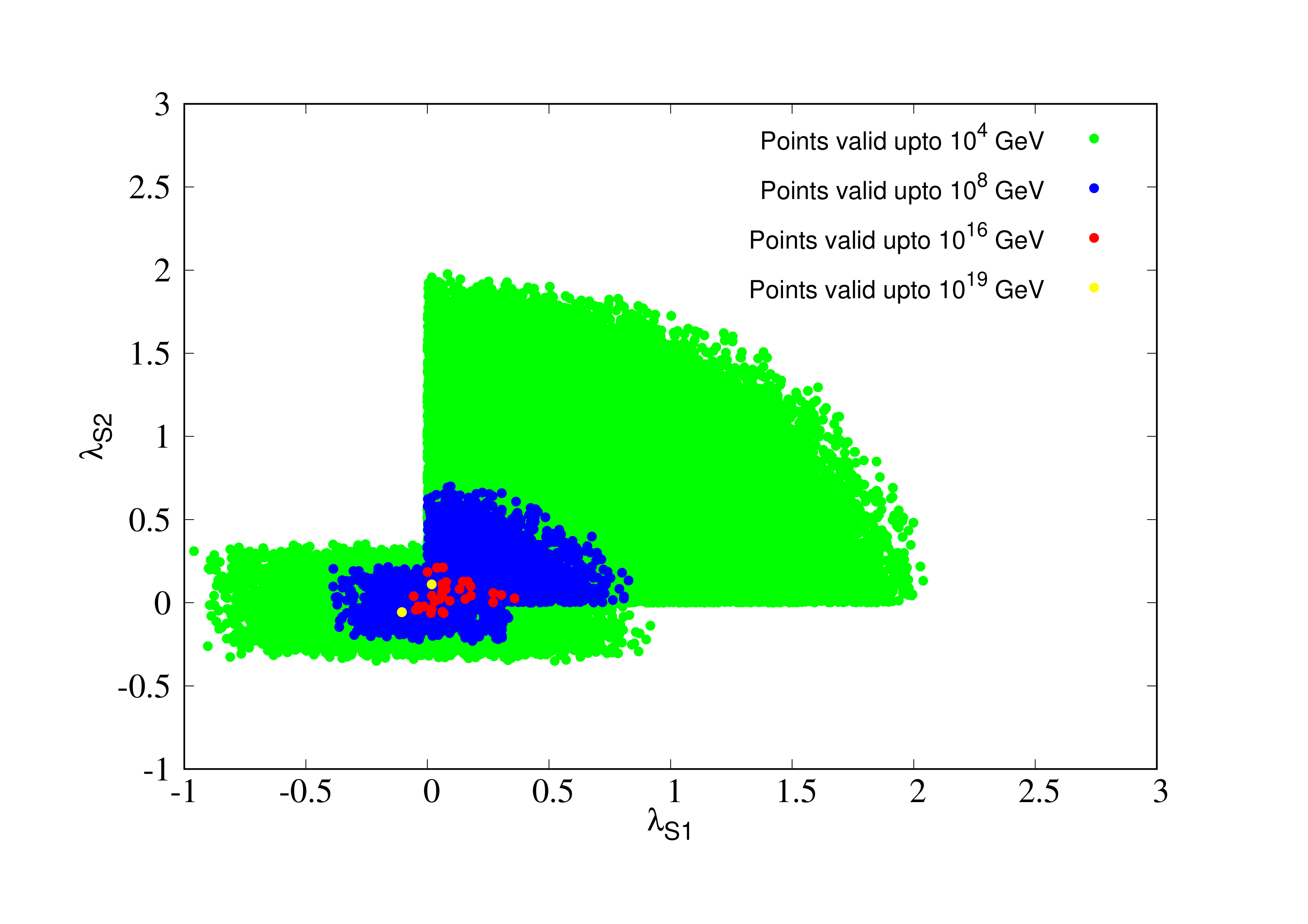}
    \caption{}\label{type2_ls1ls2}
    \end{subfigure} %
    \qquad
    \begin{subfigure}{.44\linewidth}
    \centering
    \includegraphics[width=8.0cm, height=6.0cm]{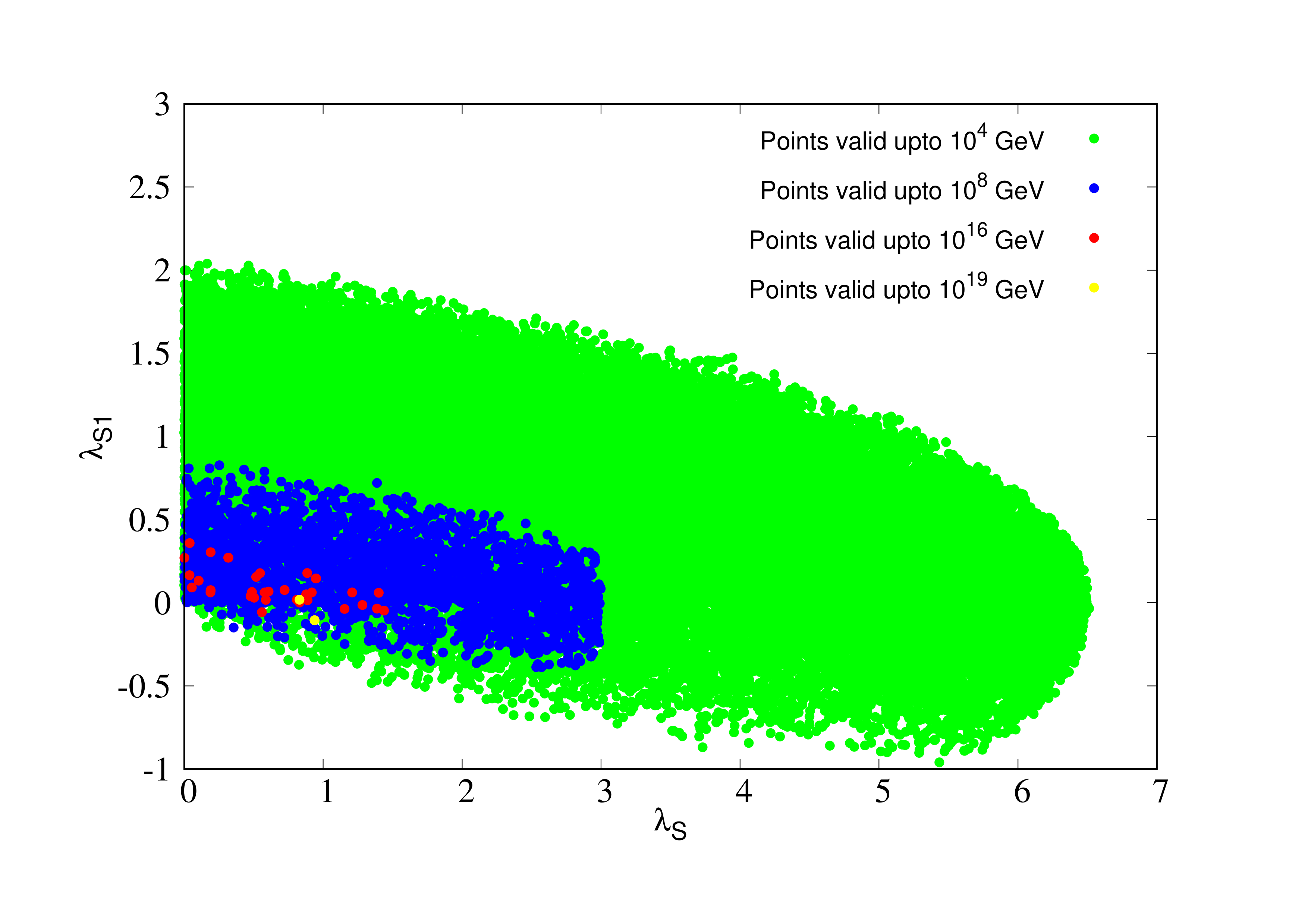}
    \caption{}\label{type2_ls1ls}
   \end{subfigure}
\\[2ex]
    \begin{subfigure}{.44\linewidth}
    \centering
    \includegraphics[width=8.0cm, height=6.0cm]{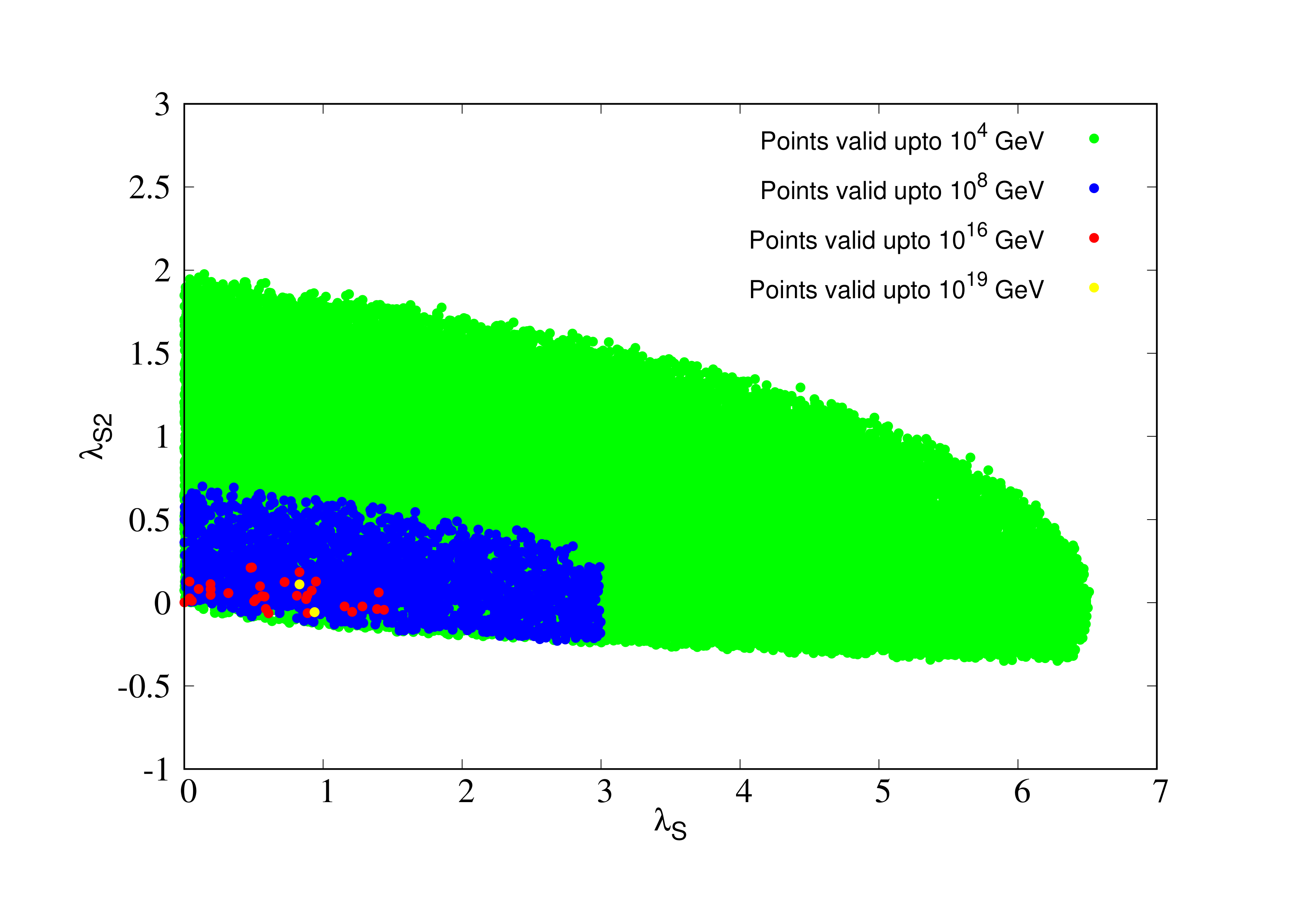}
    \caption{}\label{fig:type2_ls2ls}
    \end{subfigure} %
\RawCaption{\caption{\it The parameter space spanned by (a) $\lambda_{S1}-\lambda_{S2}$ (b) $\lambda_{S}-\lambda_{S1}$ and $\lambda_{S}-\lambda_{S2}$, valid upto different high scales in Type-II 2HDM+DM scenario.}
\label{lsls1ls2_type2}}
\end{figure}

\begin{figure}[!hptb]
\floatsetup[subfigure]{captionskip=10pt}
    \begin{subfigure}{.44\linewidth}
    \centering
    \includegraphics[width=8.0cm, height=6.0cm]{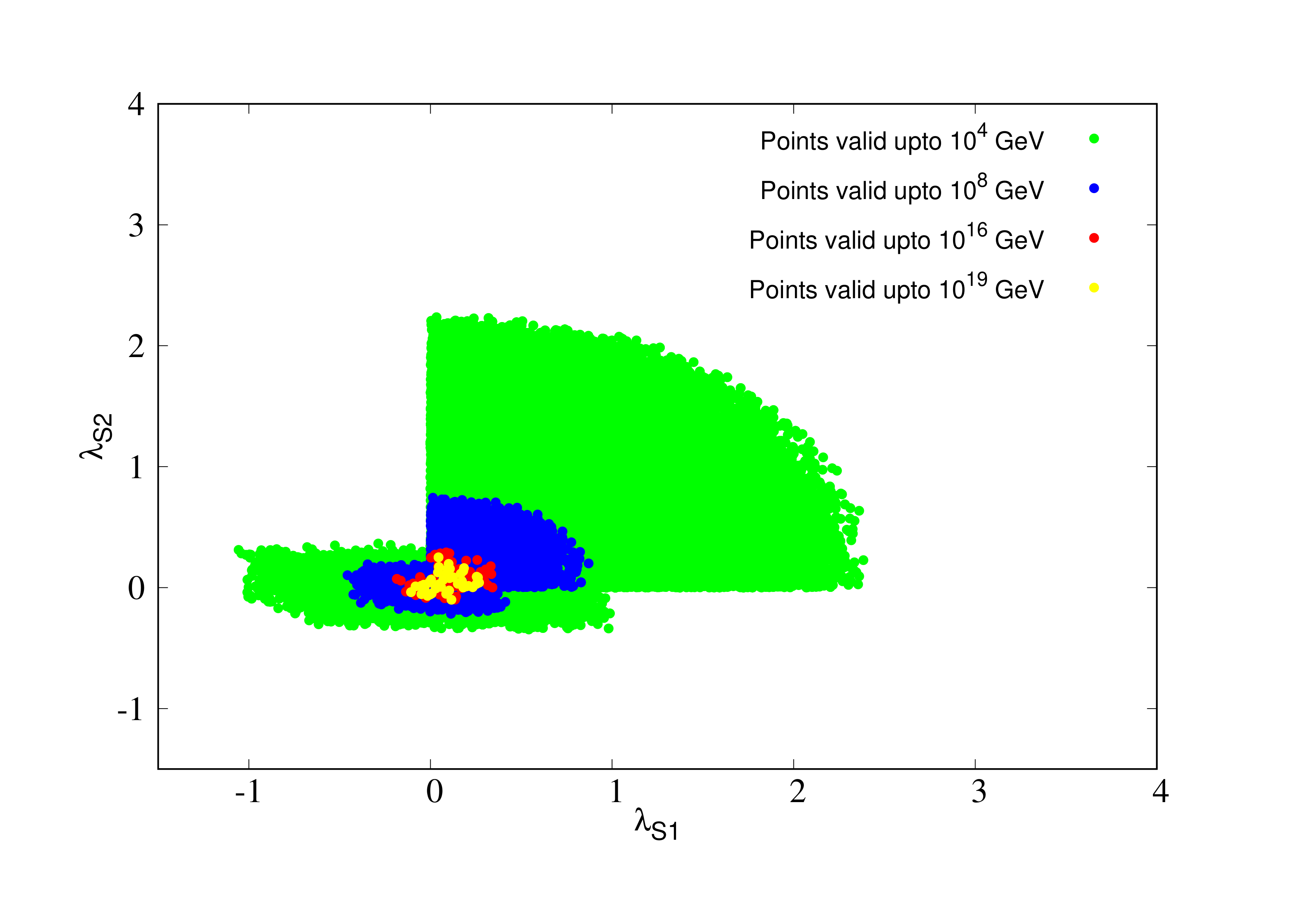}
    \caption{}\label{type2_ls1ls}
    \end{subfigure} %
    \qquad
    \begin{subfigure}{.44\linewidth}
    \centering
    \includegraphics[width=8.0cm, height=6.0cm]{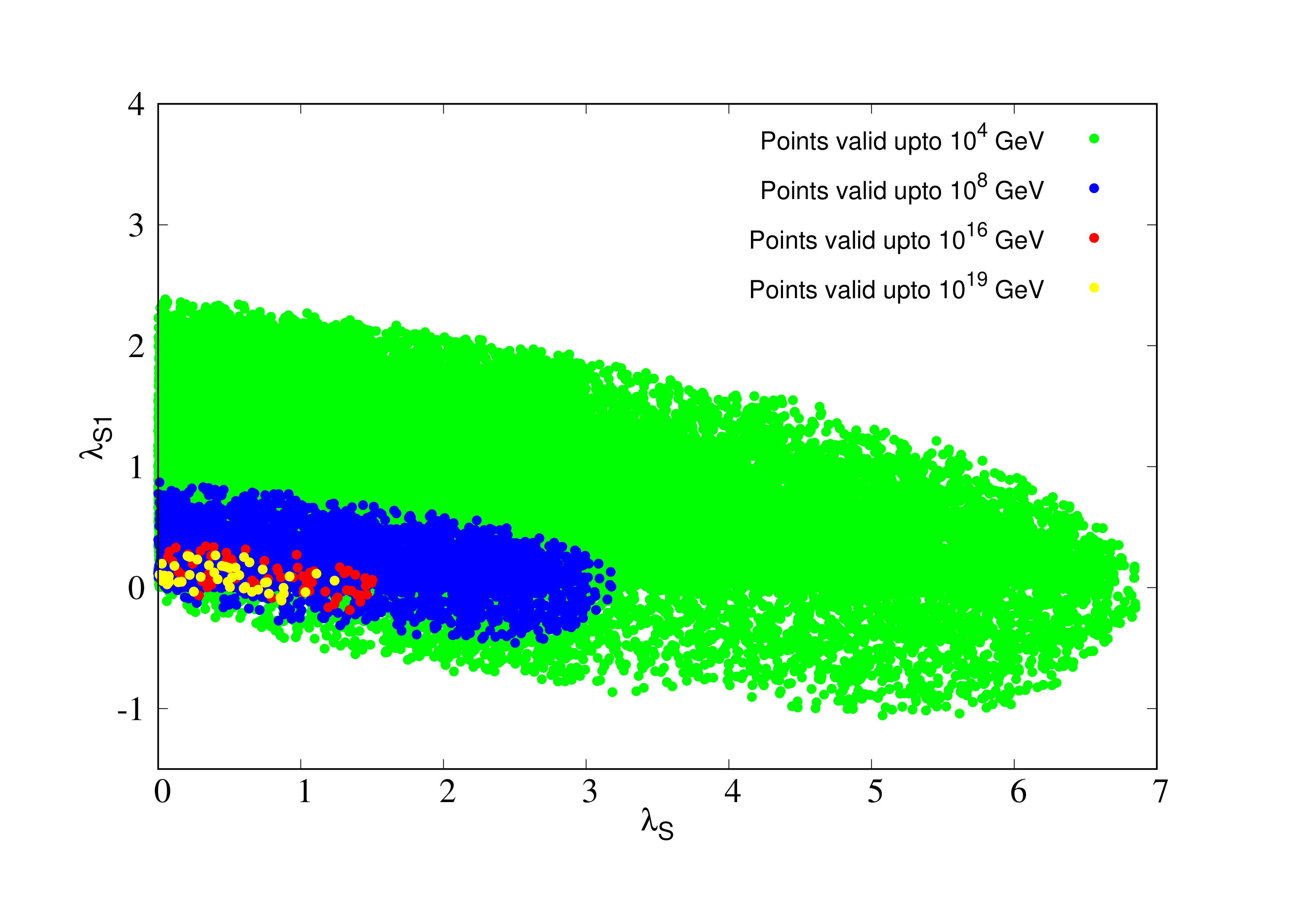}
    \caption{}\label{typex_ls1ls}
   \end{subfigure}
\\[2ex]
    \begin{subfigure}{.44\linewidth}
    \centering
    \includegraphics[width=8.0cm, height=6.0cm]{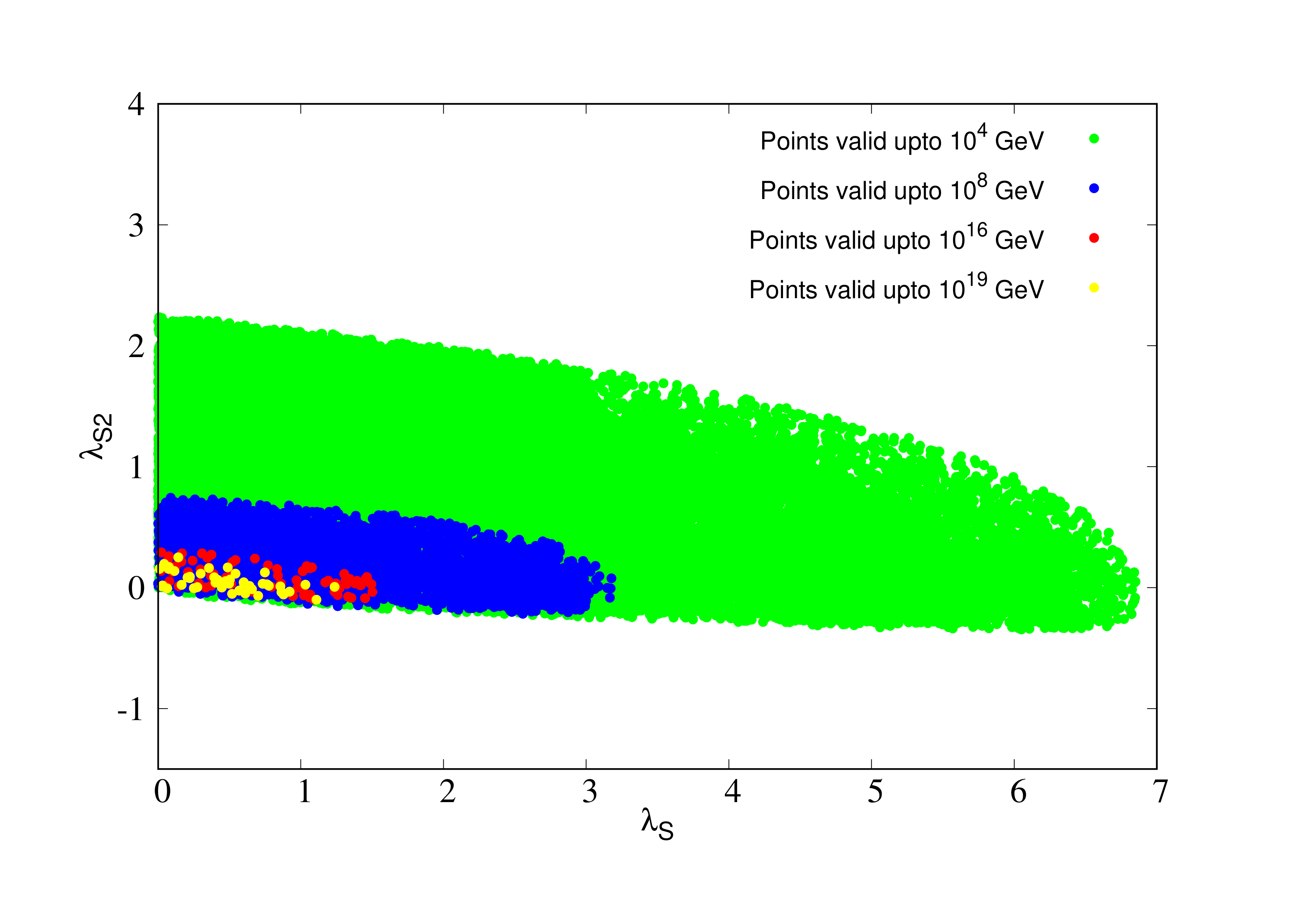}
    \caption{}\label{fig:typex_ls2ls}
    \end{subfigure} %
\RawCaption{\caption{\it The parameter space spanned by (a) $\lambda_{S1}-\lambda_{S2}$ (b) $\lambda_{S}-\lambda_{S1}$ and $\lambda_{S}-\lambda_{S2}$, valid upto different high scales in Type-X 2HDM+DM scenario.}
\label{lsls1ls2_typex}}
\end{figure}

In Figure~\ref{lsls1ls2_type2} and \ref{lsls1ls2_typex}, we show the parameter space valid upto various high-scales in case of Type-II+singlet DM and Type-X+singlet DM scenarios respectively. 
In any plots, where any two of the three couplings $\lambda_{S1}$, $\lambda_{S2}$ and $\lambda_S$ are shown, the third remaining coupling has been varied from $-4\pi$ to $4\pi$ in the scatter plots. Similar marginalization has been carried out for the remaining parameters in the scalar potential.

 In Figure~\ref{lsls1ls2_type2}(a),(b) and (c), we can see that the $\lambda_{S1}$, $\lambda_{S2} \lsim 2.0, 0.8, 0.3$ and $0.1$ and $\lambda_{S} \lsim 6.5, 3.0, 1.5$ and $1$ in order for the model to be valid upto 10 TeV, $10^8$ GeV, $10^{16}$ GeV and $10^{20}$ GeV respectively in Type-II. The results are very similar in case of Type-X as can be seen in Figure~\ref{lsls1ls2_typex}(a),(b) and (c). In that case, $\lambda_{S1}$, $\lambda_{S2} \lsim 2.4, 0.9, 0.4$ and $0.3$ and $\lambda_{S} \lsim 6.8, 3.2, 1.7$ and $1.2$ in order for the model to be valid upto 10 TeV, $10^8$ GeV, $10^{16}$ GeV and $10^{20}$ GeV respectively. 

The aforementioned constraints come largely from perturbative unitarity. As we have discussed earlier, the $\lambda_S$ coupling runs the fastest among all scalar couplings and therefore the perturbativity is driven by $\lambda_S$. If we see Equations~\ref{lambda_rge_type2} and \ref{lambda_rge_typex}, we see that the running of $\lambda_S$ depends strongly on $\lambda_{S1}$ as well as $\lambda_{S2}$. It is also clear that when $\lambda_{S1}$ and $\lambda_{S2}$ are positive, increasing $\lambda_{S}$ will imply stronger limits on $\lambda_{S1}$ as well as $\lambda_{S2}$. This is also clear from Figure~\ref{lsls1ls2_type2}(b),(c) as well as \ref{lsls1ls2_typex}(b),(c) .
In Figures~\ref{lsls1ls2_type2} and \ref{lsls1ls2_typex}, the regions where either of $\lambda_{S1}$ and $\lambda_{S2}$ is negative, get constrained by the requirement of vacuum stability (Equations~\ref{stability}) as well. 

We would like to point out that, if we compare the regions in Figures~\ref{lsls1ls2_type2} and ~\ref{lsls1ls2_typex}, valid upto various scales, we see that the regions follow very similar pattern in Type-II and Type-X, although the Yukawa sectors in both cases are different. The ranges of high scale validity are comparable in both cases, though the allowed region is slightly bigger in Type-X compared to Type-II. We have further checked that, if we are phenomenologically allowed to start with exactly same low-scale values of all the parameters in Type-II and Type-X, the high scale upto which the theories will be valid differs by $\lsim 100$ GeV. The comparison between the two scenarios in this respect can be summarized as follows: 

\begin{itemize}
\item As long as $\lambda_S$ is moderate to large, the perturbativity constraints are strongly driven by $\lambda_S$. The running of $\lambda_S$ has very little contribution from the Yukawa sector, since the Yukawa coupling-dependent terms enter in the running of $\lambda_S$, indirectly via $\lambda_{S1}$ and $\lambda_{S2}$. Moreover, Yukawa sector differs for Type-II and Type-X only in terms of $Y_b$, which is a small quantity compared to the other terms in the running. Therefore, when the $\lambda_S$ plays dominant role in high-scale validity, little difference is expected between Type-II ad Type-X.
\item When $\lambda_S$ is extremely small, the perturbative unitarity is driven by the quartic couplings $\lambda_1$ or $\lambda_3$. Although the runnings of these couplings directly involve Yukawa terms, the smallness of bottom Yukawa ensures that the running remains almost same for Type-II and Type-X.
\item We have checked that for the same benchmarks, the limits of high-scale validity differ by $\lsim 100$ GeV between Type-II and Type-X 2HDM.
\item We have further explored the possible difference in the allowed parameter space Type-II and Type-X, after imposing high-scale validity as well as the experimental constraints discussed earlier. This can be seen in terms of the upper limit on $\lambda_{S1}$ and $\lambda_{S2}$. The allowed parameter space  is larger for Type-X. The reason is as follows. In Type-X the non-standard scalar masses can be low even after all the collider and B-physics constraints are applied. But in Type-II 2HDM requirements from B-physics as well as collider constraints imply large lower limits on non-standard scalar masses. This in turn necessitates large values of quartic couplings in the EW-scale (see Equations~\ref{eq:paratran}). Therefore, as a consequence of RG-running the limits on $\lambda_{S1}$ and $\lambda_{S2}$ become stronger in case of Type-II as compared to Type-X, following the requirement of perturbative unitarity of the quartic couplings. 
\item We also see that in the regions allowed upto GUT scale or Planck scale (red and yellow points) in Figures~\ref{lsls1ls2_type2} and \ref{lsls1ls2_typex}, Type-II case has much fewer points. In the Type-II the heavy scalars have to be much heavier compared to the 125 GeV Higgs (constraint coming from B-physics and collider observables), as can be seen from Equations~\ref{eq:paratran}. It is therefore extremely difficult to get quartic couplings that are small.
\end{itemize}

\subsection{Constraints from DM sector}

An important question arises: which fractions of the parameter regions discussed above are consistent with constraints on a scalar DM? 
With this in mind, we look next for parameter regions that are allowed by the relic density~\cite{Planck:2018vyg} and direct DM search experiments such as XENON~\cite{XENON:2018voc,XENON:2020kmp,XENON:2022ltv}, PANDA-X~\cite{PandaX-II:2021nsg,PandaX-4T:2021bab} and LUX-ZEPLIN~\cite{LZ:2022ufs}. In this work we have implemented our models in {\tt Feynrules}~\cite{Christensen:2008py} and calculated the DM observables with {\tt micrOMEGAs}~\cite{Belanger:2013oya}.

Let us discuss Type-II and Type-X cases one by one. Since DM mass plays an important role in DM-DM annihilation as well as DM-nucleon scattering, we present our results in both cases for three benchmark DM masses, namely $m_{\text{DM}} = 400, 200$ and 62.5 GeV. While implementing the limit from the observed relic density from PLANCK~\cite{Planck:2018vyg}, we have made sure that relic density for our parameter points do not exceed the $2\sigma$ upper bound. We have also ensured that our DM candidate accounts for at least 10\% of the total observed relic, since there is always a possibility that there are multiple DM candidates in nature which can account for the observed relic density collectively. However, we also indicate the regions of parameter space that give rise to the observed relic within $2\sigma$ uncertainty.

\begin{figure}[!hptb]
\floatsetup[subfigure]{captionskip=10pt}
    \begin{subfigure}{.44\linewidth}
    \centering
    \includegraphics[width=8.0cm, height=6.0cm]{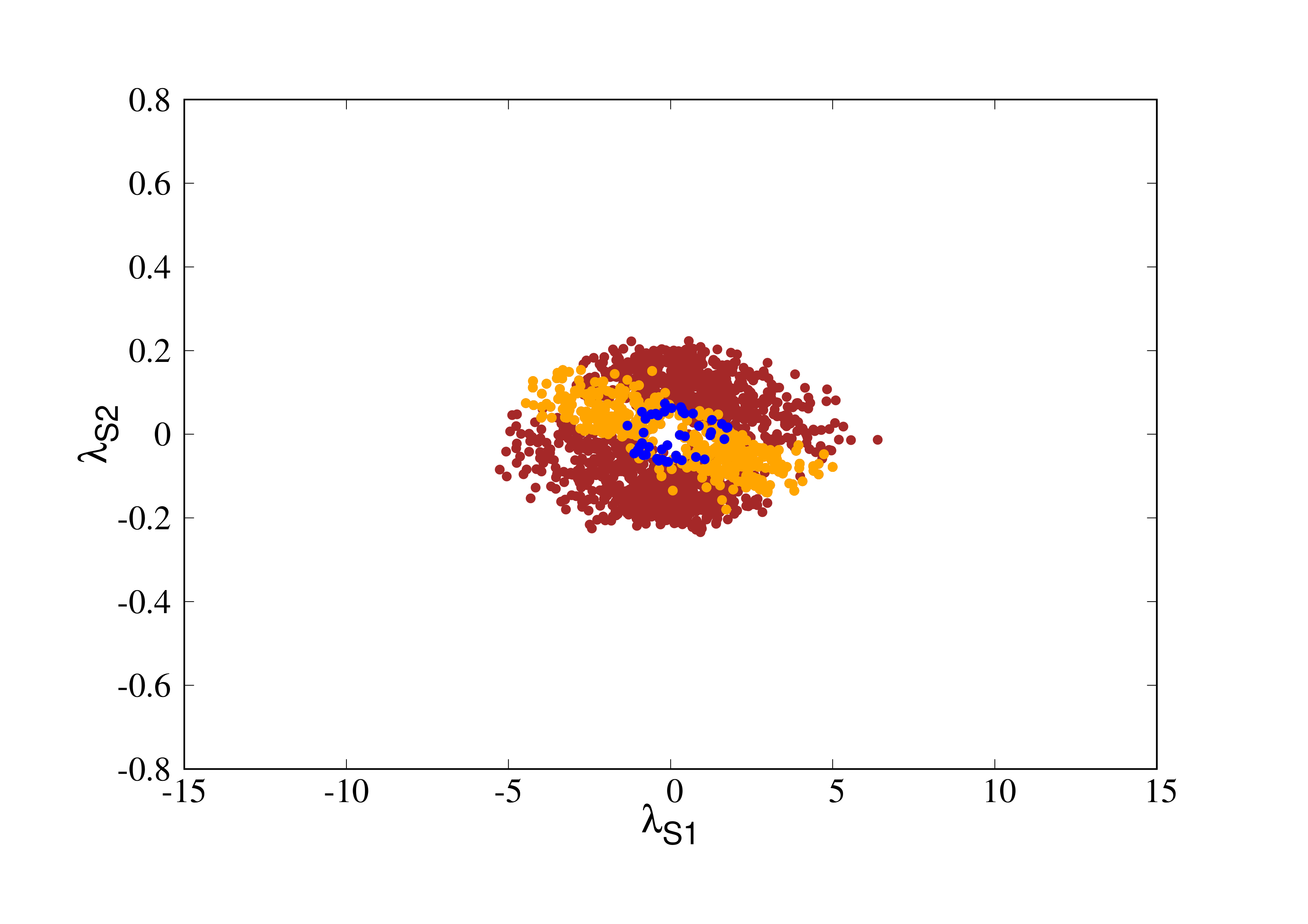}
    \caption{}\label{typex_relic_dd_400}
    \end{subfigure} %
    \qquad
    \begin{subfigure}{.44\linewidth}
    \centering
    \includegraphics[width=8.0cm, height=6.0cm]{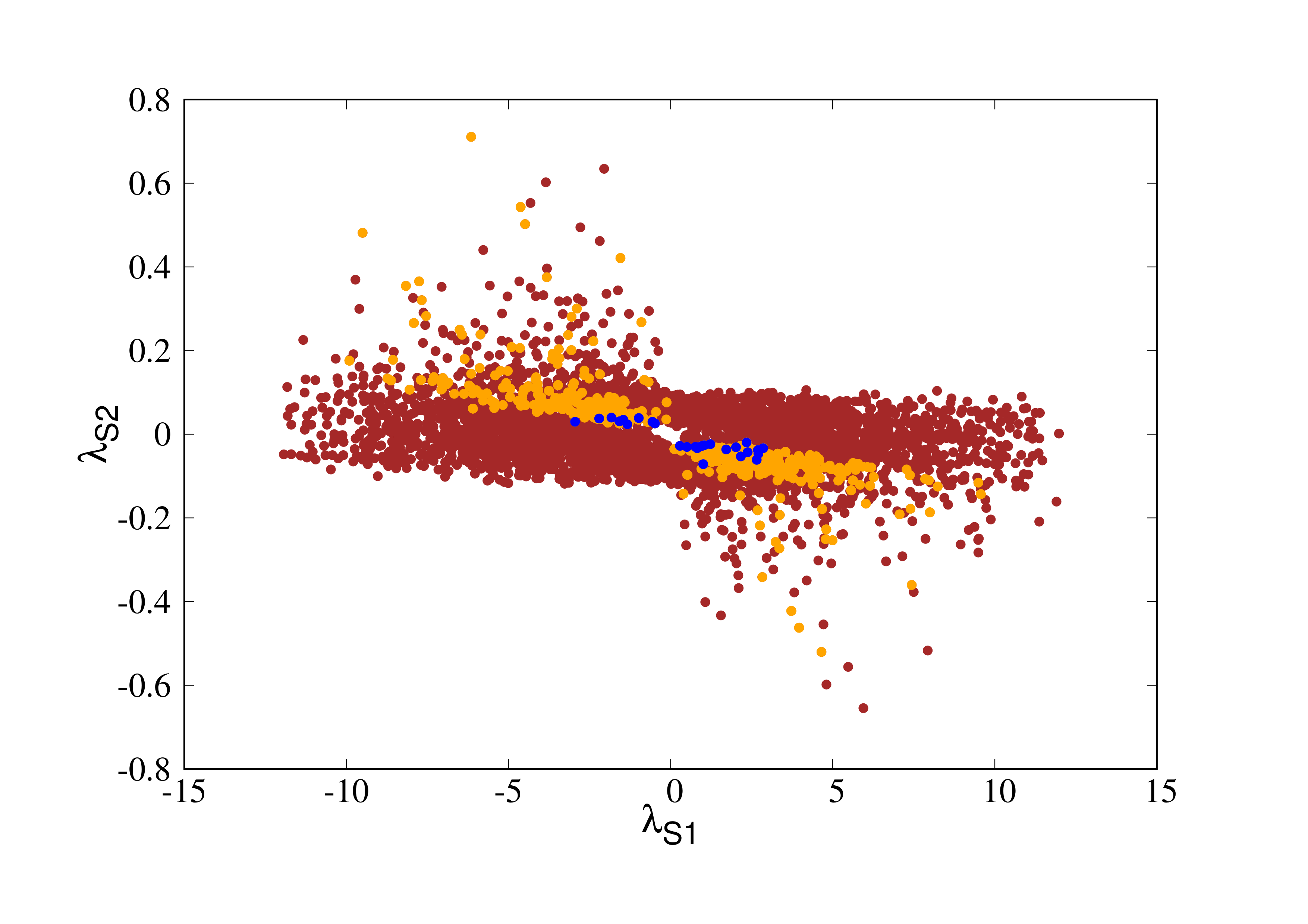}
    \caption{}\label{typex_relic_dd_200}
   \end{subfigure} \\
    \begin{subfigure}{.44\linewidth}
    \centering
    \includegraphics[width=8.0cm, height=6.0cm]{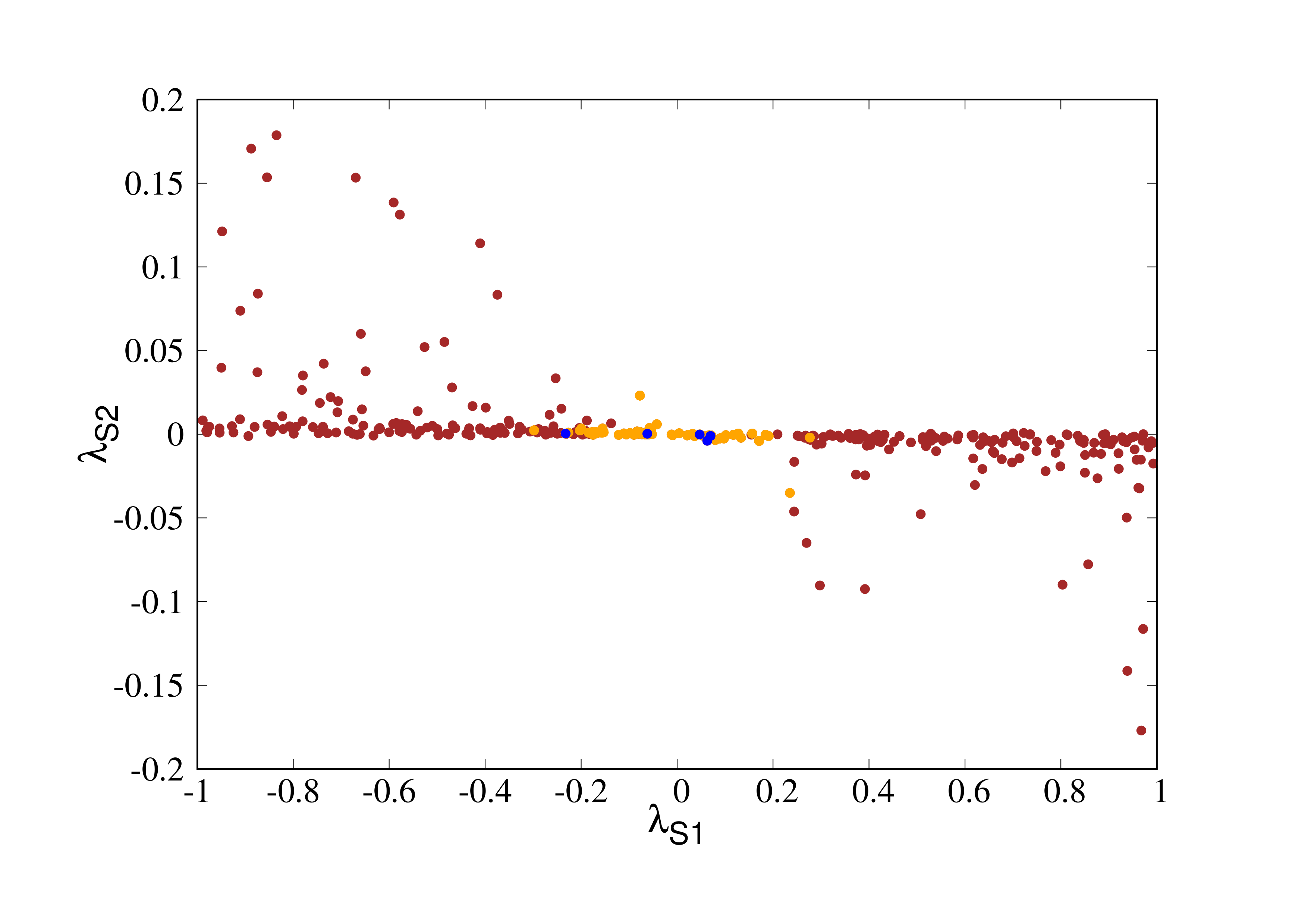}
    \caption{}\label{type2_relic_dd_62.5}
   \end{subfigure}

\RawCaption{\caption{\it The allowed parameter space in Type-II 2HDM, spanned by $\lambda_{S1}-\lambda_{S2}$ for (a) $m_{DM} = 400$ GeV and (b) $m_{DM} = 200$ GeV and (c) $m_{DM} = 62.5$ GeV . The maroon points have at least 10\% contribution to observed relic density, while the orange points satisfy direct detection constraints in addition. The white region at the centre in ~\ref{relic_dd_type2}(a) and ~\ref{relic_dd_type2}(b) are disallowed by relic over-abundance. The blue points satisfy the actual observed relic density as well as direct detection bound.}
\label{relic_dd_type2}}
\end{figure}

\begin{figure}[!hptb]
\floatsetup[subfigure]{captionskip=10pt}
    \begin{subfigure}{.44\linewidth}
    \centering
    \includegraphics[width=8.0cm, height=6.0cm]{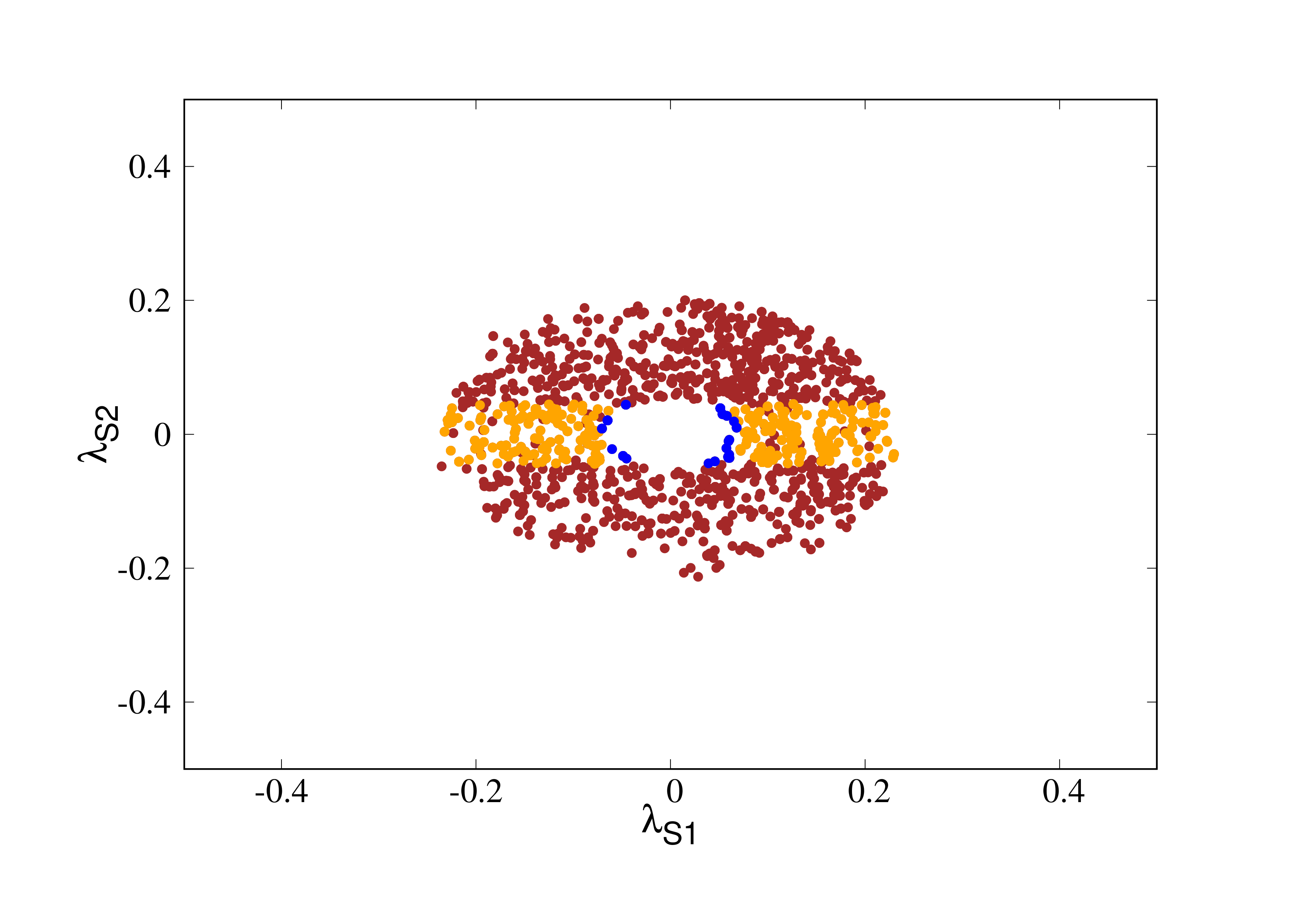}
    \caption{}\label{typex_relic_dd_400}
    \end{subfigure} %
    \qquad
    \begin{subfigure}{.44\linewidth}
    \centering
    \includegraphics[width=8.0cm, height=6.0cm]{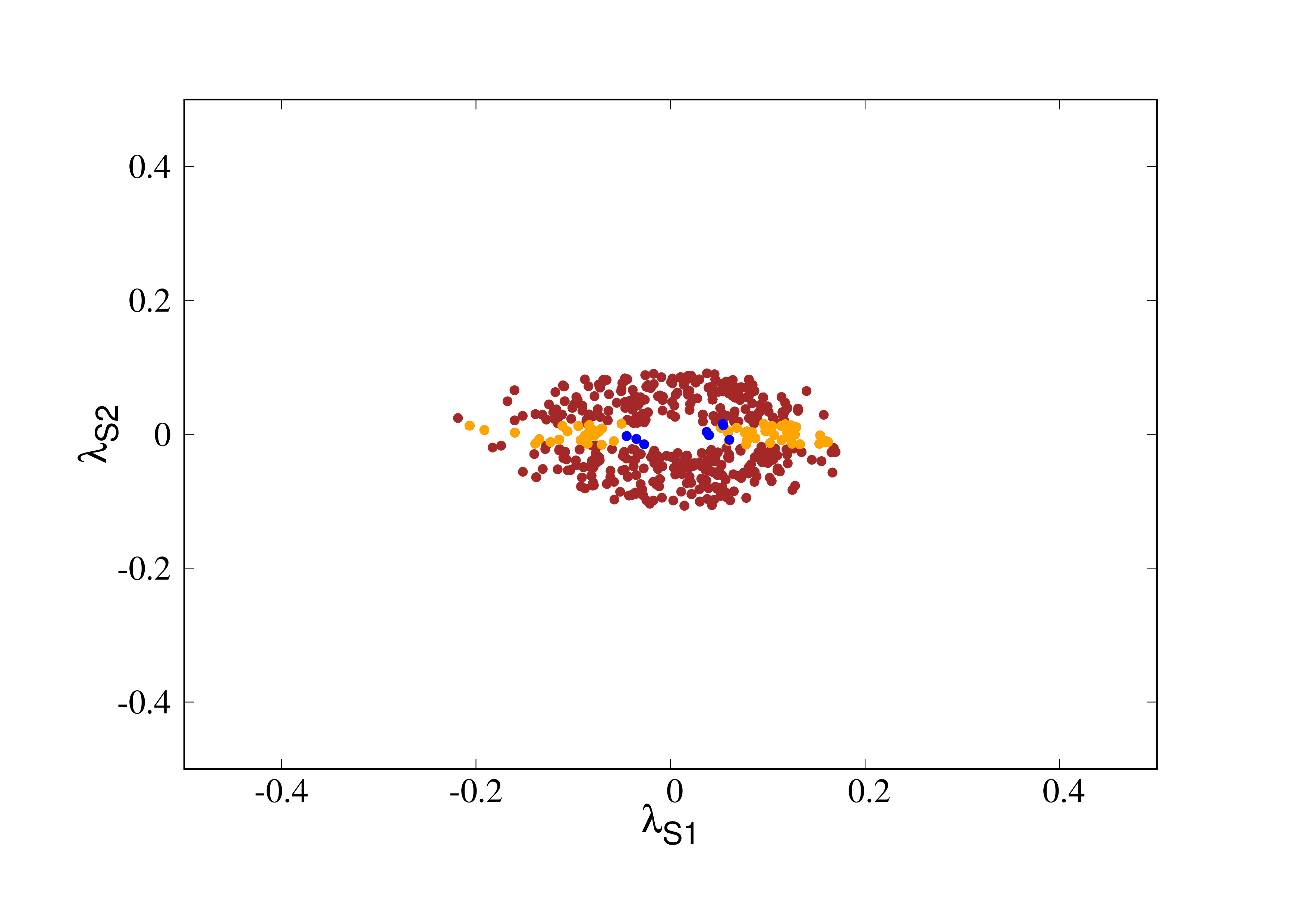}
    \caption{}\label{typex_relic_dd_200}
   \end{subfigure}\\
    \begin{subfigure}{.44\linewidth}
    \centering
    \includegraphics[width=8.0cm, height=6.0cm]{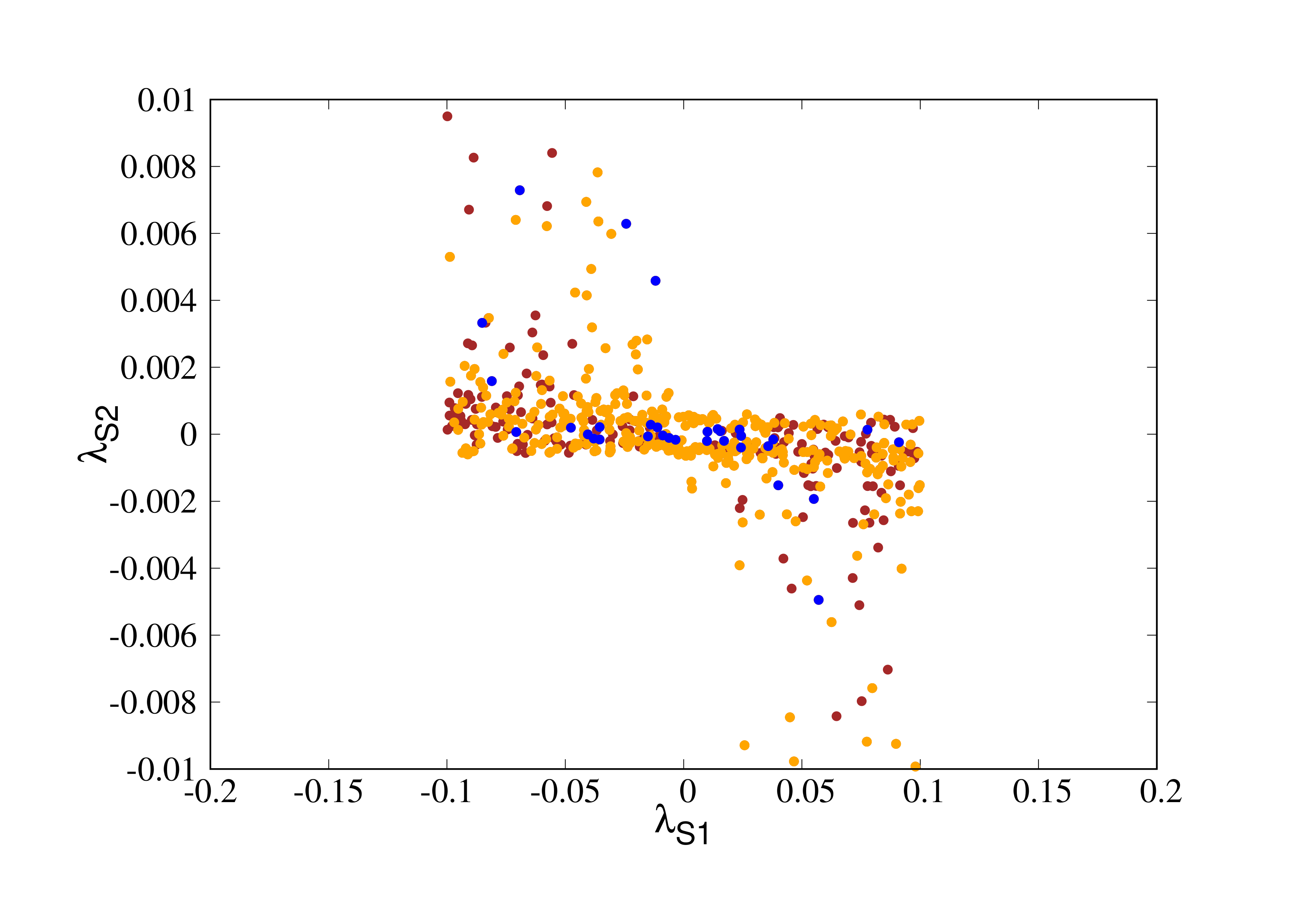}
    \caption{}\label{typex_relic_dd_62.5}
   \end{subfigure}
\\
\RawCaption{\caption{\it Same as Figure~\ref{relic_dd_type2}, but for Type-X 2HDM. }
\label{relic_dd_typex}}
\end{figure}
 
In Figure~\ref{relic_dd_type2}, we present our results for Type-II 2HDM. Figure~\ref{relic_dd_type2}(a), (b) and (c) represent the allowed parameter space for $m_{\text{DM}} = 400, 200$ and 62.5 GeV respectively, the maroon points satisfy under-relic upto 10\%, and the orange points satisfy the upper limit from direct search experiments in addition. The major annihilation channel for the DM pair in this case is into a pair of Higgses. Since in Type-II, the non-standard scalars are heavy $\gsim 600$ GeV, the kinematically favored annihilation channel is into SM-Higgs pair, which is typically governed by the $\lambda_{S2}$, coupling. Therefore, in all the plots we see the range of $\lambda_{S2}$ is restricted by the observed relic density to $|\lambda_{S2}| \lsim 0.2$(0.1) for 400 GeV (200 GeV) DM mass, whereas the limit on $\lambda_{S1}$ is more relaxed. However, the limit on $\lambda_{S1}$ in case of $m_\text{DM}=400$ GeV is stronger compared to $m_\text{DM}=200$ GeV case. The reason is, for $m_\text{DM} = 400$ GeV, an additional $Hh$ final state also opens up. Therefore, in this case, the parameter space becomes relic under-abundant with smaller $\lambda_{S1}$ compared to $m_\text{DM}=200$ GeV case. One can see a small region in the center of the $\lambda_{S1}-\lambda_{S2}$ plane, which is disallowed by the relic over-abundance. $|\lambda_{S1}| > 0.9 (1)$ for 400 GeV (200 GeV) DM mass. 

A special mention is in order for the Higgs-resonance region. Since one of the major annihilation channels for the DM pair is into $b \bar b$ final states, the DM mass in the Higgs resonance implies large annihilation cross-section and major under-abundance, unless the relevant coupling $\lambda_{S2}$ is very small, as can be seen from Figure~\ref{relic_dd_type2}(c). In this region, understandably, the dependence on $\lambda_{S1}$ is further diminished compared to the other mass points, from the point of view of relic density.

The DM-nucleon elastic scattering cross-section in Type-II 2HDM, shows an interesting pattern in the allowed parameter space. Since the coupling $\lambda_{S2}$,  which plays a crucial role in the annihilation of DM-pairs, is also responsible for the DM-nucleon scattering, a small DM-nucleon scattering cross-section will necessarily imply small annihilation cross-section and consequently large relic over-abundance. This problem can be avoided in specific regions of the parameter space especially regions of $\tan\beta$, where due to enhanced couplings of the second doublet to down-type quarks in Type-II we can have cancellation between contributions coming from $t$-channel elastic scatterings involving the two neutral scalars. This effect was pointed out in ~\cite{Dey:2019lyr} earlier. Therefore, in Figure~\ref{relic_dd_type2}, (a) and (b) we see the orange points which are allowed by both observed relic and direct search bound have specific correlation between the two couplings $\lambda_{S1}$ and $\lambda_{S2}$. For $m_\text{DM}=62.5$ GeV i.e. in the vicinity of 125 GeV Higgs resonance, we would get very strong limit on $\lambda_{S2}$ from relic under-abundance as mentioned earlier, which automatically ensures very small direct search cross-section, as well as BR($h_{\text SM} \rightarrow \text invisible) \lsim 19\%$~\cite{CMS:2018yfx}. However, in this case the upper bound from direct search does put a strong limit on $|\lambda_{S1}| \lsim 0.3$.

In Figures~\ref{relic_dd_typex}, we present the allowed parameter space for the Type-X scenario. Here too, the maroon points satisfy observed relic density (at least 10\% of observed relic) and the orange points satisfy the upper bound from direct search experiments in addition. Here too, the preferred annihilation channels are into a pair of scalars. Notably, in Type-X 2HDM, non-standard scalar masses are allowed to be low, therefore annihilation into second non-standard CP-even scalars as well as charged Higgses also takes place. Therefore, we see both $\lambda_{S1}$ and $\lambda_{S2}$ become constrained by the observed relic in this case.  We can see both $|\lambda_{S1}|$ and $|\lambda_{S2}|$, $\lsim 0.2 (0.15)$ $m_\text{DM}=400 (200)$ GeV from relic under-abundance. 
On the other hand, we see a small region in the middle of the $\lambda_{S1}$-$\lambda_{S2}$ plane disallowed from over-abundance. The corresponding limits on $|\lambda_{S1}|, |\lambda_{S2}| \gsim 0.1 (0.08)$ for 400 GeV (200 GeV) DM mass.
When DM mass is in the vicinity of 125 GeV Higgs resonance, the coupling $|\lambda_{S2}|$ becomes strongly restricted from relic under-abundance, to $\lsim 0.01$  whereas the other coupling $\lambda_{S1}$ is naturally less constrained and can vary upto $|\lambda_{S1}| \lsim 0.1$. One will see the reverse behavior in terms of the two couplings if the DM mass is in the vicinity of resonance of the non-standard scalar.

The region allowed by the direct search experiments shows a very different pattern in Type-X case compared to Type-II 2HDM. Since here the only coupling participating in the DM-nucleon elastic scattering is $\lambda_{S2}$ (the quarks couple to $\Phi_2$ in this case), the upper bound from direct search experiments only constrains $\lambda_{S2}$, while keeping $\lambda_{S1}$ completely free, as can be seen from Figures~\ref{relic_dd_typex}. When $m_\text{DM}$ is in the vicinity of Higgs resonance, the smallness of $\lambda_{S2}$ demanded by relic under-abundance, necessarily ensures small direct search cross-section, similar to Type-II case. However, unlike Type-II, here the direct search bound does not constrain $\lambda_{S1}$ at all.

One should note that, for our chosen DM mass range of few hundred GeV, the presence of light non-standard scalars make the annihilation process stronger in Type-X model as compared to Type-II. A more restrictive outer contour for Type-X, compared to that in Type-II (see Figures~\ref{relic_dd_type2} and \ref{relic_dd_typex}) is a result of demanding at least 10\% of the observed relic in both cases. Similarly, the constraint from relic over-abundance affecting the central regions of the plots is more relaxed in Type-X compared to Type-II.   

We would like to remind here that the scalar singlet having masses of the order of hundred GeV is still allowed by relic density and direct search constraints having two portal interactions with SM-like and heavy/light Higgs, unlike the usual Higgs-portal scenario, where only SM Higgs portal is present. We also note that while calculating direct search constraints for the relic under-abundant situation (upto 10\% contribution), we have been conservative to take the full direct search cross-section, without folding it by the appropriate factor. We have also checked that the parameter space considered here satisfies the constraints from indirect detection.

\subsection{Combining high-scale validity with DM constraints}

Having illustrated the regions allowed by perturbative unitarity, vacuum stability upto various high scales, and after studying the regions allowed by DM constraints, namely, observed relic density and direct search, we will confront the two types of constraints with each other. We have seen in Figures~\ref{lsls1ls2_type2} and \ref{lsls1ls2_typex} that the high scale validity pushes the DM-portal couplings ($\lambda_{S1}$ and $\lambda_{S2}$) as well as DM self-coupling ($\lambda_S$) to smaller values. We have pointed out the upper limits on these couplings, previously. We also point out regions allowed by DM constraints, namely observed relic and direct search. Both these constraints affect only the DM-portal couplings $\lambda_{S1}$ and $\lambda_{S2}$. In Figures~\ref{relic_dd_type2} and ~\ref{relic_dd_typex}, we saw very small couplings are disfavored since they will overclose the universe. Interestingly, this is in tension with the high-scale validity discussed before. In order to examine the contrast between the two, we present a comparison between the two competing constraints in Table~\ref{type2_dm_highscale} for Type-II and \ref{typex_dm_highscale} for Type-X. We have seen in the previous section that the upper and lower limits on $\lambda_{S1}$ and $\lambda_{S2}$ vary a little bit with DM mass. However, it was also evident that the limits from relic over-abundance did not change substantially between 400 GeV and 200 GeV DM mass. Therefore, in order to avoid confusion, we quote only single numbers in the last row Table~\ref{type2_dm_highscale} and ~\ref{typex_dm_highscale}. Although these numbers can vary slightly with DM mass, our major conclusion remains.

\begin{table}[!hptb]
\begin{center}
\begin{tabular}{|c|c|c|}
\hline
Constraint &  Limits on $\lambda_{S1}$ & Limit on $\lambda_{S2}$ \\
\hline
High-scale validity upto $10^{6}$ GeV & $\lambda_{S1} \lsim 1.0$ & $\lambda_{S2} \lsim 1.0$ \\
\hline
High-scale validity upto $10^{8}$ GeV & $\lambda_{S1} \lsim 0.8$ & $\lambda_{S2} \lsim 0.8$ \\
\hline
High-scale validity upto $10^{16}$ GeV & $\lambda_{S1} \lsim 0.3$ & $\lambda_{S2} \lsim 0.3$ \\
\hline
High-scale validity upto $10^{19}$ GeV & $\lambda_{S1} \lsim 0.1$ & $\lambda_{S2} \lsim 0.1$ \\
\hline
\hline
DM constraints & $\lambda_{S1} \gsim 1.0$  & $\lambda_{S2}  \gsim 0.1$ \\
\hline
\end{tabular}
\caption{Type-II: Regions allowed by vacuum stability and perturbative unitarity upto different high scales and regions allowed by DM constraints namely observed relic density as well as direct search of DM. These limits are relaxed in various mass-resonance regions, which is discussed below.}
\label{type2_dm_highscale}
\end{center}
\end{table}

\begin{table}[!hptb]
\begin{center}
\begin{tabular}{|c|c|c|}
\hline
Constraint &  Limits on $\lambda_{S1}$ & Limit on $\lambda_{S2}$ \\
\hline
High-scale validity upto $10^{8}$ GeV & $\lambda_{S1} \lsim 0.9$ & $\lambda_{S2} \lsim 0.9$ \\
\hline
High-scale validity upto $10^{16}$ GeV & $\lambda_{S1} \lsim 0.4$ & $\lambda_{S2} \lsim 0.4$ \\
\hline
High-scale validity upto $10^{19}$ GeV & $\lambda_{S1} \lsim 0.3$ & $\lambda_{S2} \lsim 0.3$ \\
\hline
\hline
DM constraints & $\lambda_{S1} \gsim 0.1$  & $\lambda_{S2}  \gsim 0.1$ \\
\hline
\end{tabular}
\caption{Type-X: Regions allowed by vacuum stability and perturbative unitarity upto different high scales and regions allowed by DM constraints namely observed relic density as well as direct search of DM. Similar relaxations as in the Type-II case are also applicable here. }
\label{typex_dm_highscale}
\end{center}
\end{table}

\noindent
It is clear from the Table~\ref{type2_dm_highscale}, that in Type-II+DM scenario, the parameter space can be valid upto $\sim 10^{6}$ GeV and not higher, in order to remain consistent with the DM constraints. On the other hand, as we see in Table~\ref{typex_dm_highscale}, the DM constraints in case of Type-X are less restrictive, and completely satisfy the requirements for validity upto very high scales. To be precise, the entire region of the parameter space consistent with DM constraints is also valid upto the Planck scale in case of Type-X 2HDM in association with a real singlet scalar DM. 
We would like to highlight this as an important contrast between Type-II and Type-X scenarios in the presence of real singlet scalar DM.

In the discussion and interpretation of Tables~\ref{type2_dm_highscale} and ~\ref{typex_dm_highscale}, a few comments are in order. The aforementioned limits are strictly valid away from the resonance regions. For example, in Figures~\ref{relic_dd_type2}(c) and \ref{relic_dd_typex}(c), we see that even with extremely small couplings $\lambda_{S1}$ and $\lambda_{S2}$, the constraints from relic density are satisfied, since the s-channel Higgs-mediated annihilation cross-section is large in this region ($m_{\text{DM}} \approx \frac{m_h}{2}$). Similar relaxation of the relic density constraints occurs when DM-mass is in the vicinity of the CP-even heavy Higgs resonance i.e. $m_{\text{DM}} \approx \frac{m_H}{2}$. In the resonance regions, for both Type-II and Type-X models, DM-constraints allowed parameter spaces can be valid upto the Planck scale.

We also note that the relic density and direct search constraints allow the models to have a cut-off as high as $10^6$ GeV or higher, so that the freeze-out of the DM (with $x\sim 20$) is not affected for DM masses upto TeV scale, with dominant depletion contribution coming after EWSB.

\subsection{Prospects at the LHC}

We briefly comment on the prospect of probing the regions of our models, which are valid upto high scales as well as consistent with DM constraints, at the LHC. In Type-II, we have seen that simultaneous satisfaction of both types of constraints leads to a maximum admissible validity scale $10^{6}$ GeV. The DM-portal couplings $\lambda_{S1}$ and $\lambda_{S2}$ are in this case $\gsim 1$. In our earlier work~\cite{Dey:2019lyr} we have seen that this region ($\lambda_{S1}, \lambda_{S2} \gsim 1$) can be probed at the high-luminosity LHC (3000 $fb^{-1}$) with $\sim 3 \sigma$ significance with cut-based analysis. This happens particularly when a non-standard scalar is produced in vector boson fusion and then decays into a DM pair. It has been shown in that work that the corresponding gluon fusion channel performs rather poorly in this case. Further improvement is possible using machine-learning techniques, as pointed out in~\cite{Dey:2019lyr}. In Type-X, although smaller $\lambda_{S1}, \lambda_{S2}$ couplings are allowed from high-scale validity as well as DM constraints as we pointed out in the previous subsection, there too, at least $\lambda_{S1}, \lambda_{S2} \gsim 1$ couplings are required to probe it in the high-luminosity LHC with 3000$fb^{-1}$ data~\cite{Dey:2021alu}. 
Therefore, although the type-X scenario can be valid upto as much as the Planck scale, even after imposing DM constraints, the regions that can be probed at the LHC are restricted to validity limits around $10^8$ GeV.

\section{Summary and Conclusions}
\label{sec5}

\noindent
We have explored the high-scale validity in terms of perturbativity, unitarity and vacuum stability, of two-Higgs doublet models with a real singlet scalar DM candidate. Such an exploration can be expected to yield useful guidelines on scenarios with an extended Higgs sector as the DM portal In this context, we have considered Type-II 2HDM, which derives its motivation from supersymmetry and Type-X 2HDM, that allows for a low mass pseudo-scalar and provides at least a partial solution to the observed ($g_{\mu}-2$) anomaly. After obtaining the one- and two-loop RG running equations with appropriate modifications/extensions in {\tt SARAH} and {\tt 2HDME}, we have identified the differences between the two aforementioned scenarios, in terms of high-scale behavior. 

We applied all the experimental constraints on both the models. The B-physics observables as well as direct collider search experiments push the lower limit for non-standard scalars to much higher values in Type-II, as compared to Type-X. The presence of the low mass pseudo-scalars in Type-X 2HDM, not only contributes to $g_{\mu}-2$, but also allows for much smaller quartic couplings, at the electro-weak scale, as compared to the Type-II scenario. This in turn, after RG-running, leads to larger allowed regions of parameter space upto various high scales in Type-X case. We have also compared the high-scale validity of 2HDM+DM scenario, with normal 2HDM cases, which was analysed in~\cite{Dey:2021pyn}. We see that the high-scale validity is generally worsened in the presence of a real-singlet DM. 

We further study the impact of the high-scale validity on the DM sector. The existing constraints, namely the observed relic density and upper bound from direct search experiments put limits on the portal couplings between DM and the scalar sector. The high scale validity of the model crucially relies on the DM constraints, as the perturbative unitarity of the portal couplings often governs the cut-off scale. In this work, we have identified the regions of parameter space of the aforementioned models, that are allowed by the DM constraints, and are also valid upto various high scales. We find that the Type-II 2HDM+real singlet DM scenario can only be valid upto $\sim 10^{6}$ GeV from the requirement of perturbative unitarity and vacuum stability, while obeying all the DM constraints at the same time. This implies, such a scenario will require intervention of new physics around $\sim 10^{6}$ GeV, in order to be viable from the standpoint of particle phenomenology as well as DM-related observations. In Type-X 2HDM + real singlet DM, on the other hand, the restrictions are much more relaxed because of the less stringent phenomenological constraints on the parameter space. It can be valid upto Planck scale while at the same time being allowed by all the existing DM search results. Finally, we comment on the discovery prospect at the high-luminosity LHC, of the regions of the parameter space in these models, that are valid upto high scales, and are also allowed by DM constraints. We find that Type-II 2HDM+real singlet DM, which is valid upto $\sim 10^6$ GeV, can be probed at the high-luminosity LHC. On the other hand, although its Type-X counterpart can be valid upto the Planck scale, only the portion of its parameter space, valid upto $\sim 10^8$ GeV can be probed at the high-luminosity LHC.

\section{Acknowledgements}

AD and JL would like to thank Indian Institute of Science Education and Research, Kolkata, where part of the work was done. 

\newpage

\begin{appendix}

\section{Two-loop RGE's}
\label{2looprge}

Here we listed the two-loop RGE's of gauge, Yukawa and quartic couplings for our scenario.
\subsection{Type-II}
\begin{align}
\label{gauge_rge_2loop}
(16 \pi ^2 \beta _{g_1})_{2-loop} &=(16 \pi ^2 \beta _{g_1})^{2-loop}_{2HDM}, \nonumber \\
(16 \pi ^2 \beta _{g_2})_{2-loop} &=(16 \pi ^2 \beta _{g_2})^{2-loop}_{2HDM}, \nonumber \\
(16 \pi ^2 \beta _{g_3})_{2-loop} &=(16 \pi ^2 \beta _{g_3})^{2-loop}_{2HDM}.
\end{align}
The $(16 \pi ^2 \beta _{g_i})^{2-loop}_{2HDM}$, for $i= 1, 2, 3$ represent the RGE's of gauge couplings for general 2HDM's at two-loop level and can be found in~\cite{Chowdhury:2015yja}.
 \begin{align}
(16 \pi ^2 \beta _{Y_t})_{2-loop} &=(16 \pi ^2 \beta _{Y_t})^{2-loop}_{2HDM} + \frac{\lambda_{S2}^2 Y_t}{16 \pi^2}, \nonumber \\ 
(16 \pi ^2 \beta _{Y_b})_{2-loop}&=(16 \pi ^2 \beta _{Y_b})^{2-loop}_{2HDM} + \frac{\lambda_{S1}^2 Y_b}{16 \pi^2}, \nonumber \\
(16 \pi ^2 \beta _{Y_{\tau}})_{2-loop} &=(16 \pi ^2 \beta _{Y_{\tau}})^{2-loop}_{2HDM} + \frac{\lambda_{S1}^2 Y_{\tau}}{16 \pi^2}.
\label{yuk_rge_2loop}
\end{align}
Here too, $(16 \pi ^2 \beta _{Y_j})^{2-loop}_{2HDM}$, where $j$ can be $t, b$ or $\tau$, are the two-loop RGE's for general 2HDM's (can be different for different types). One can easily find out the structure of them from~\cite{Chowdhury:2015yja}. Next is the RGE's of quartic couplings for our model.
\begin{align}
(16 \pi ^2 \beta _{\lambda_1})_{2-loop} &=(16 \pi ^2 \beta _{\lambda_1})^{2-loop}_{2HDM} + 4\lambda_{S1}^2 +\frac{ (-32\lambda_{S1}^3 -20 \lambda_1 \lambda_{S1}^2)}{16 \pi^2},  \nonumber \\ 
(16 \pi ^2 \beta _{\lambda_2})_{2-loop}&=(16 \pi ^2 \beta _{\lambda_2})^{2-loop}_{2HDM} + 4\lambda_{S2}^2 +\frac{ (-32\lambda_{S2}^3 -20 \lambda_2 \lambda_{S2}^2)}{16 \pi^2}, \nonumber \\
(16 \pi ^2 \beta _{\lambda_3})_{2-loop} &=(16 \pi ^2 \beta _{\lambda_3})^{2-loop}_{2HDM} + 4\lambda_{S1}\lambda_{S2} +\frac{( -16(\lambda_{S1}^2 \lambda_{S2}+\lambda_{S2}^2 \lambda_{S1}) -2 \lambda_3(\lambda_{S1}^2+\lambda_{S2}^2+ 8 \lambda_{S1}\lambda_{S2}))}{16 \pi^2}, \nonumber \\ 
(16 \pi ^2 \beta _{\lambda_4})_{2-loop}&=(16 \pi ^2 \beta _{\lambda_4})^{2-loop}_{2HDM} + \frac{(-2 \lambda_4(\lambda_{S1}^2+\lambda_{S2}^2+ 8 \lambda_{S1}\lambda_{S2}))}{16 \pi ^2}, \nonumber \\
(16 \pi ^2 \beta _{\lambda_5})_{2-loop} &=(16 \pi ^2 \beta _{\lambda_5})^{2-loop}_{2HDM} + \frac{(-2 \lambda_5(\lambda_{S1}^2+\lambda_{S2}^2+ 8 \lambda_{S1}\lambda_{S2}))}{16 \pi ^2}.
\end{align}

\begin{align}
(16 \pi ^2 \beta _{\lambda_S})_{2-loop} &=(16 \pi ^2 \beta _{\lambda_S})^{1-loop}+ \frac{288}{3} g^2_1 (\lambda_{S1}^2 + \lambda_{S2}^2) + 288 g^2_2 (\lambda_{S1}^2 + \lambda_{S2}^2) - 384 (\lambda_{S1}^3 + \lambda_{S2}^3) \nonumber \\
&- 80 \lambda_S (\lambda_{S1}^2 + \lambda_{S2}^2)- \frac{17}{3} \lambda_{S}^3 - 288 \lambda_{S1}^2 Y^2_{b} - 96 \lambda_{S1}^2 Y^2_{\tau} - 288 \lambda_{S2}^2 Y^2_t, \nonumber \\
(16 \pi ^2 \beta _{\lambda_{S1}})_{2-loop} &= (16 \pi ^2 \beta _{\lambda_{S1}})^{1-loop} + \frac{5}{2}g^4_1 \lambda_{S2} + \frac{15}{2}g^4_2 \lambda_{S2} + \frac{1737}{144}g^4_1 \lambda_{S1} + \frac{15}{8}g^2_1 g^2_2 \lambda_{S1} - \frac{123}{16}g^4_2 \lambda_{S1}  \nonumber \\
&- 8 \lambda_{S2}^2 \lambda_{S1} + (12 g^2_1 + 36 g^2_2) \lambda_2 \lambda_{S1} - 15 \lambda_1^2 \lambda_{S1} + (2 g^2_1 + 6 g^2_2)  \lambda_{S1}^2  - 72 \lambda_1 \lambda_{S2}^2 \nonumber \\
&+ (8 g^2_1 + 24 g^2_2) \lambda_3 \lambda_{S2} - 42 \lambda_{S1}^3 - 16 \lambda_3 \lambda_{S2}^2 - 32 \lambda_3 \lambda_{S1} \lambda_{S2} - (8 \lambda_{S2} + 2 \lambda_{S1}) \lambda_3^2 \nonumber \\
&+ (4 g^2_1 + 12 g^2_2) \lambda_4 \lambda_{S2} -16 \lambda_4 \lambda_{S1} \lambda_{S2} -(8 \lambda_{S2} + 2 \lambda_{S1})\lambda_3 \lambda_4 - (8 \lambda_{S2} + 2 \lambda_{S1}) \lambda_4^2 - 12 \lambda_{S1}^2 \lambda_5 \nonumber \\
&- (12 \lambda_{S2} - \frac{13}{6} \lambda_{S1}) \lambda_5^2 - 12 (2\lambda_3+\lambda_4)\lambda_{S1}Y^2_t - 12 \lambda_1 \lambda_{S1} Y^2_{\tau} -\frac{9}{2} \lambda_{S1} Y^2_b Y^2_t \nonumber \\
& - (\frac{9}{2} Y^4_{\tau} +\frac{27}{2}Y^4_b) \lambda_{S1} -8 \lambda_{S1}^2 Y^2_{\tau} + (\frac{25}{4} g^2_1 + \frac{15}{4} g^2_2) \lambda_{S1} Y^2_{\tau}  \nonumber \\
&+ (\frac{25}{12} g^2_1 + \frac{45}{4} g^2_2 + 40 g^2_3 -24 \lambda_{S1}- 36 \lambda_1)\lambda_{S1} Y^2_b, \nonumber \\
(16 \pi ^2 \beta _{\lambda_{S2}})_{2-loop} &= (16 \pi ^2 \beta _{\lambda_{S2}})^{1-loop} + \frac{5}{2}g^4_1 \lambda_{S1} + \frac{15}{2}g^4_2 \lambda_{S1} + \frac{1737}{144}g^4_1 \lambda_{S2} + \frac{15}{8}g^2_1 g^2_2 \lambda_{S2} - \frac{123}{16}g^4_2 \lambda_{S2} \nonumber \\
&- 8 \lambda_{S1}^2 \lambda_{S2} + (12 g^2_1 + 36 g^2_2) \lambda_2 \lambda_{S2} - 15 \lambda_2^2 \lambda_{S2} + (2 g^2_1 + 6 g^2_2)  \lambda_{S2}^2  - 72 \lambda_2 \lambda_{S2}^2 \nonumber \\
&+ (8 g^2_1 + 24 g^2_2) \lambda_3 \lambda_{S1} - 42 \lambda_{S2}^3 - 16 \lambda_3 \lambda_{S1}^2 - 32 \lambda_3 \lambda_{S1} \lambda_{S2} - (8 \lambda_{S1} + 2 \lambda_{S2}) \lambda_3^2 \nonumber \\
&+ (4 g^2_1 + 12 g^2_2) \lambda_4 \lambda_{S1} -16 \lambda_4 \lambda_{S1} \lambda_{S2} -(8 \lambda_{S1} + 2 \lambda_{S2})\lambda_3 \lambda_4 - (8 \lambda_{S1} + 2 \lambda_{S2}) \lambda_4^2 - 12 \lambda_{S2}^2 \lambda_5 \nonumber \\
&- (12 \lambda_{S1} - \frac{13}{6} \lambda_{S2}) \lambda_5^2 - 12 (2\lambda_3+\lambda_4)\lambda_{S1}Y^2_b - (8 \lambda_3 + 4 \lambda_4) \lambda_{S1} Y^2_{\tau} -\frac{9}{2} \lambda_{S2} Y^2_b Y^2_t - \frac{27}{2} \lambda_{S2} Y^4_t \nonumber \\
&+ (\frac{85}{12} g^2_1 + \frac{45}{4} g^2_2 + 40 g^2_3 -24 \lambda_{S2}- 36 \lambda_2)\lambda_{S2} Y^2_t.
\label{yuk_rge_2loop}
\end{align}

\subsection{Type-X}

\begin{align}
\label{gauge_rge_2loop}
(16 \pi ^2 \beta _{g_1})_{2-loop} &=(16 \pi ^2 \beta _{g_1})^{2-loop}_{2HDM}, \nonumber \\
(16 \pi ^2 \beta _{g_2})_{2-loop} &=(16 \pi ^2 \beta _{g_2})^{2-loop}_{2HDM}, \nonumber \\
(16 \pi ^2 \beta _{g_3})_{2-loop} &=(16 \pi ^2 \beta _{g_3})^{2-loop}_{2HDM}.
\end{align}
The $(16 \pi ^2 \beta _{g_i})^{2-loop}_{2HDM}$, for $i= 1, 2, 3$ represent the RGE's of gauge couplings for general 2HDM's at two-loop level and can be found in~\cite{Chowdhury:2015yja}.
 \begin{align}
(16 \pi ^2 \beta _{Y_t})_{2-loop} &=(16 \pi ^2 \beta _{Y_t})^{2-loop}_{2HDM} + \frac{\lambda_{S2}^2 Y_t}{16 \pi^2}, \nonumber \\ 
(16 \pi ^2 \beta _{Y_b})_{2-loop}&=(16 \pi ^2 \beta _{Y_b})^{2-loop}_{2HDM} + \frac{\lambda_{S2}^2 Y_b}{16 \pi^2}, \nonumber \\
(16 \pi ^2 \beta _{Y_{\tau}})_{2-loop} &=(16 \pi ^2 \beta _{Y_{\tau}})^{2-loop}_{2HDM} + \frac{\lambda_{S1}^2 Y_{\tau}}{16 \pi^2}.
\label{yuk_rge_2loop}
\end{align}
Here too, $(16 \pi ^2 \beta _{Y_j})^{2-loop}_{2HDM}$, where $j$ can be $t, b$ or $\tau$, are the two-loop RGE's for general 2HDM's (can be different for different types). One can easily find out the structure of them from~\cite{Chowdhury:2015yja}. Next is the RGE's of quartic couplings for our model.
\begin{align}
(16 \pi ^2 \beta _{\lambda_1})_{2-loop} &=(16 \pi ^2 \beta _{\lambda_1})^{2-loop}_{2HDM} + 4\lambda_{S1}^2 +\frac{ (-32\lambda_{S1}^3 -20 \lambda_1 \lambda_{S1}^2)}{16 \pi^2},  \nonumber \\ 
(16 \pi ^2 \beta _{\lambda_2})_{2-loop}&=(16 \pi ^2 \beta _{\lambda_2})^{2-loop}_{2HDM} + 4\lambda_{S2}^2 +\frac{ (-32\lambda_{S2}^3 -20 \lambda_2 \lambda_{S2}^2)}{16 \pi^2}, \nonumber \\
(16 \pi ^2 \beta _{\lambda_3})_{2-loop} &=(16 \pi ^2 \beta _{\lambda_3})^{2-loop}_{2HDM} + 4\lambda_{S1}\lambda_{S2} +\frac{( -16(\lambda_{S1}^2 \lambda_{S2}+\lambda_{S2}^2 \lambda_{S1}) -2 \lambda_3(\lambda_{S1}^2+\lambda_{S2}^2+ 8 \lambda_{S1}\lambda_{S2}))}{16 \pi^2}, \nonumber \\ 
(16 \pi ^2 \beta _{\lambda_4})_{2-loop}&=(16 \pi ^2 \beta _{\lambda_4})^{2-loop}_{2HDM} + \frac{(-2 \lambda_4(\lambda_{S1}^2+\lambda_{S2}^2+ 8 \lambda_{S1}\lambda_{S2}))}{16 \pi ^2}, \nonumber \\
(16 \pi ^2 \beta _{\lambda_5})_{2-loop} &=(16 \pi ^2 \beta _{\lambda_5})^{2-loop}_{2HDM} + \frac{(-2 \lambda_5(\lambda_{S1}^2+\lambda_{S2}^2+ 8 \lambda_{S1}\lambda_{S2}))}{16 \pi ^2}.
\end{align}

\begin{align}
(16 \pi ^2 \beta _{\lambda_S})_{2-loop} &=(16 \pi ^2 \beta _{\lambda_S})^{1-loop}+ \frac{288}{3} g^2_1 (\lambda_{S1}^2 + \lambda_{S2}^2) + 288 g^2_2 (\lambda_{S1}^2 + \lambda_{S2}^2) - 384 (\lambda_{S1}^3 + \lambda_{S2}^3) \nonumber \\
&- 80 \lambda_S (\lambda_{S1}^2 + \lambda_{S2}^2)- \frac{17}{3} \lambda_{S}^3 - 288 \lambda_{S2}^2 Y^2_{b} - 96 \lambda_{S1}^2 Y^2_{\tau} - 288 \lambda_{S2}^2 Y^2_t, \nonumber \\
(16 \pi ^2 \beta _{\lambda_{S1}})_{2-loop} &= (16 \pi ^2 \beta _{\lambda_{S1}})^{1-loop} + \frac{5}{2}g^4_1 \lambda_{S2} + \frac{15}{2}g^4_2 \lambda_{S2} + \frac{1737}{144}g^4_1 \lambda_{S1} + \frac{15}{8}g^2_1 g^2_2 \lambda_{S1} - \frac{123}{16}g^4_2 \lambda_{S1}  \nonumber \\
&- 8 \lambda_{S2}^2 \lambda_{S1} + (12 g^2_1 + 36 g^2_2) \lambda_2 \lambda_{S1} - 15 \lambda_1^2 \lambda_{S1} + (2 g^2_1 + 6 g^2_2)  \lambda_{S1}^2  - 72 \lambda_1 \lambda_{S2}^2 \nonumber \\
&+ (8 g^2_1 + 24 g^2_2) \lambda_3 \lambda_{S2} - 42 \lambda_{S1}^3 - 16 \lambda_3 \lambda_{S2}^2 - 32 \lambda_3 \lambda_{S1} \lambda_{S2} - (8 \lambda_{S2} + 2 \lambda_{S1}) \lambda_3^2 \nonumber \\
&+ (4 g^2_1 + 12 g^2_2) \lambda_4 \lambda_{S2} -16 \lambda_4 \lambda_{S1} \lambda_{S2} - (8 \lambda_{S2} + 2 \lambda_{S1})\lambda_3 \lambda_4 - (8 \lambda_{S2} + 2 \lambda_{S1}) \lambda_4^2 - 12 \lambda_{S1}^2 \lambda_5 \nonumber \\
&- (12 \lambda_{S2} - \frac{13}{6} \lambda_{S1}) \lambda_5^2 - 12 (2\lambda_3+\lambda_4)\lambda_{S2}(Y^2_t + Y^2_b) - 12 \lambda_1 \lambda_{S1} Y^2_{\tau} - \frac{9}{2} Y^4_{\tau} \lambda_{S1}-8 \lambda_{S1}^2 Y^2_{\tau} \nonumber \\
&+ (\frac{25}{4} g^2_1 + \frac{15}{4} g^2_2) \lambda_{S1} Y^2_{\tau},  \nonumber \\
(16 \pi ^2 \beta _{\lambda_{S2}})_{2-loop} &= (16 \pi ^2 \beta _{\lambda_{S2}})^{1-loop} + \frac{5}{2}g^4_1 \lambda_{S1} + \frac{15}{2}g^4_2 \lambda_{S1} + \frac{1737}{144}g^4_1 \lambda_{S2} + \frac{15}{8}g^2_1 g^2_2 \lambda_{S2} - \frac{123}{16}g^4_2 \lambda_{S2}  \nonumber \\
&- 8 \lambda_{S1}^2 \lambda_{S2} + (12 g^2_1 + 36 g^2_2) \lambda_2 \lambda_{S2} - 15 \lambda_2^2 \lambda_{S2} + (2 g^2_1 + 6 g^2_2)  \lambda_{S2}^2  - 72 \lambda_2 \lambda_{S2}^2 \nonumber \\
&+ (8 g^2_1 + 24 g^2_2) \lambda_3 \lambda_{S1} - 42 \lambda_{S2}^3 - 16 \lambda_3 \lambda_{S1}^2 - 32 \lambda_3 \lambda_{S1} \lambda_{S2} - (8 \lambda_{S1} + 2 \lambda_{S2}) \lambda_3^2 \nonumber \\
&+ (4 g^2_1 + 12 g^2_2) \lambda_4 \lambda_{S1} -16 \lambda_4 \lambda_{S1} \lambda_{S2} - (8 \lambda_{S1} + 2 \lambda_{S2})\lambda_3 \lambda_4 - (8 \lambda_{S1} + 2 \lambda_{S2}) \lambda_4^2 - 12 \lambda_{S2}^2 \lambda_5 \nonumber \\
&- (12 \lambda_{S1} - \frac{13}{6} \lambda_{S2}) \lambda_5^2 - 4 (2\lambda_3+\lambda_4)\lambda_{S1}Y^2_{\tau} - frac{27}{2}(Y^4_t + Y^4_b) \lambda_{S2} - 21 \lambda_{S2} Y^2_b Y^2_t \nonumber \\
&- (36 \lambda_2 + 24 \lambda_{S2}) \lambda_{S2} Y^2_t + (\frac{25}{12} g^2_1 + \frac{45}{4} g^2_2 + 40 g^2_3 -24 \lambda_{S2}- 36 \lambda_2)\lambda_{S2} Y^2_b \nonumber \\
& + (\frac{85}{12} g^2_1 + \frac{45}{4} g^2_2 + 40 g^2_3)\lambda_{S2} Y^2_t.
\label{yuk_rge_2loop}
\end{align}

\noindent
In the case of the usual quartic couplings present in general 2HDM's, which are $\lambda_{1,..5}$, we use the term $(16 \pi ^2 \beta _{\lambda_k})^{2-loop}_{2HDM}$($k=1,..5$) to represent the two-loop RGE's of respective couplings at two-loop in general 2HDM model for different types (see~\cite{Chowdhury:2015yja}). On the other hand, for the other three quartic couplings, namely $\lambda_S, \lambda_{S1}$ and $\lambda_{S2}$, the terms, $(16 \pi ^2 \beta _{\lambda_l})^{1-loop}$ ($l=S, S1$ or $S2$) represent the one-loop RGE's of respective coupling for our case for Type-II and Type-X 2HDM respectively.

\end{appendix}

\bibliographystyle{JHEP}
\bibliography{ref1}
\end{document}